\documentclass[10pt]{aa} 
\usepackage{graphics}
\usepackage{natbib}
\bibpunct{(}{)}{;}{}{}{,}
\begin{document} 
 
   \thesaurus{3          
	      (
	       02.18.7;  
	       09.04.1;  
	       11.09.1 NGC 6946;
               11.19.2;  
               11.19.6;  
	       13.09.1   
	       )}

   \title{Monte Carlo Predictions of Far-Infrared Emission from Spiral Galaxies}
 
 
   \author{
   S. Bianchi\inst{1,2}
   \and J. I.\,Davies\inst{1} \and P. B.\,Alton\inst{1} }
 
   \offprints{bianchi@mpia-hd.mpg.de}
 
   \institute{ Department of Physics \& Astronomy, Cardiff University,
   PO Box 913, Cardiff CF2 3YB, Wales, U.K. 
   \and Max-Planck-Institute f\"ur Astronomie, K\"onigstuhl 17, D-69117  
   Heidelberg, Germany.
   } 
 
   \date{Received 6 March 2000/ Accepted 5 May 2000} 
 
   \maketitle 
 
   \begin{abstract} 

We present simulations of Far Infrared (FIR) emission by dust in spiral
galaxies, based on the Monte Carlo radiative transfer code of
\citet*{BianchiApJ1996}. The radiative transfer is carried out at several
wavelength in the Ultraviolet (UV), optical and Near Infrared (NIR), to cover 
the range of the stellar Spectral Energy Distribution (SED). Together with
the images of the galactic model, a map of the energy absorbed by dust
is produced. Using Galactic dust properties, the spatial distribution of
dust temperature is derived under the assumption of thermal equilibrium.
A correction is applied for non-equilibrium emission in the Mid Infrared
(MIR). Images of dust emission can then be produced at any wavelength in
the FIR.

We show the application of the model to the spiral galaxy NGC~6946. The
observed stellar SED is used as input and models are produced for
different star-dust geometries. It is found that only optically thick
dust disks can reproduce the observed amount of FIR radiation. However,
it is not possible to reproduce the large FIR scalelength suggested 
by recent observation of spirals at 200$\mu$m, even when the scalelength 
of the dust disk is larger than that for stars. Optically thin models
have ratios of optical/FIR scalelengths closer to the 200$\mu$m observations, 
but with smaller absolute scalelengths than optically thick cases.
The modelled temperature distributions are compatible with observations
of the Galaxy and other spirals. We finally discuss the approximations of 
the model and the impact of a clumpy stellar and dust structure on the
FIR simulations.


\end{abstract} 

\section{Introduction}

Dark lanes in edge-on spirals clearly reveal the presence of dust in 
galactic disks. Depending on the dust amount and distribution, the
extinction of starlight may affect to different degrees our knowledge of
the objects harboring dust and of the distant universe in the
background. However, deriving the amount of dust from the extinction
features in spiral galaxies is not easy, because of the uncertainty in
the relative distributions of dust and stars. Indeed, realistic
models are needed to avoid equivocal results \citep{DisneyMNRAS1989}.
Recently, \citet{XilourisSub1998} have analised a sample of seven
edge-on spirals by fitting optical and NIR images with a complex
radiative transfer model inclusive of scattering. They derived a
small opacity, with their exponential dust disks having a mean face-on 
optical depth in the B-band $\tau_\mathrm{B}=0.8$. Unfortunately, the 
method can only be applied successfully to a limited numer of cases, 
i.e. edge-on galaxies, where it is possible to analyze the vertical 
surface-brightness distribution and the dust extinction is maximised by 
the inclination.

Further constrains on the structure and opacity of dust disks can be
obtained by studying dust emission. Assuming
that all the stellar radiation absorbed by dust is reemitted in the MIR
and (mainly) in the FIR, the amount of extinction in a galaxy can be simply 
derived by comparing the total dust emission with the stellar one (the {\em
energy balance} technique). The extinction is then used, together
with a radiative transfer model, to study the dust disk properties
\citep{EvansThesis1992,XuA&A1995,TrewhellaMNRAS1998}. However, the
dust emission is treated as a bulk and generally only the disk optical 
depth is derived. The parameters for the geometrical distributions of 
dust and stars have to be assumed a priori. 

Three-dimensional models of FIR emission from spiral galaxies can help
to retrieve information about the relative star-dust geometry from the 
SED of dust emission and the spatial distribution of the radiation, when
available. However, an additional factor is introduced in such models.
Not only it is necessary to assume a dust distribution, but also a
knowledge of the variation of the ISRF heating throughout the disk is
needed. The ISRF could be in principle derived from the observed
profiles of stellar emission, but a complicate de-projection is needed
involving assumption on the dust distribution. This approach is adopted
by \citet{WalterbosApJ1996}. They compute fluxes and surface brightness
gradients at 60$\mu$m and 100$\mu$m for a sample of 20 spirals. The
diffuse ISRF is derived from B band profiles and the dust distribution 
from the HI column density. The problem of defining an ISRF can be
avoided by assuming a temperature gradient for dust grains. 
\citet{DaviesMNRAS1997} modelled the 140$\mu$m and 240$\mu$m Galactic 
emission observed by DIRBE, by adopting a gradient that matches the 
longitudinal variation of temperature observed in the data themselves.

Otherwise, the dust heating by the ISRF can be derived in a self-consistent 
way from a radiative transfer model of the stellar radiation in the 
dusty environment. \citet{SautyA&A1998} produced a Monte Carlo radiative 
transfer model for the clumpy ISM of the spiral galaxy NGC~6946,
although a self-consistent ISRF is derived only for the UV radiation,
relying on optical profiles for radiation at longer wavelelngth. Dust heating 
in a clumpy galactic ISM is also included in the photometric evolution
model of \citet{SilvaApJprep1998}. They carry out the radiative transfer
in a proper way for stellar emission embedded in the clumps and use an
approximation for the smooth medium. In the emission model of 
\citet{WolfA&A1998} the radiative transfer is computed with the Monte
Carlo technique for the whole spectral range, both in the smooth medium
and clumps. However, the adopted geometry is typical of star-formation
environment.

In this paper, we present a self-consistent galactic model for the FIR 
emission. The model is based on the \citet*[][hereafter, BFG]{BianchiApJ1996} 
Monte Carlo code for the radiative transfer (complete with scattering) in 
dusty spiral galaxies. A map of the energy absorbed by dust is derived
from the output of the radiative transfer for stellar radiation at 
several wavelength. The grain temperature along the dust distribution is
therefore computed from the radiative transfer itself. Luminosity, spectrum 
and surface brightness distribution of the FIR radiation can be retrieved, 
for any wavelength. The main advantage with respect to other models of FIR
emission lies in the relative simplicity. We have limited the number of
parameters involved to a minimum, adopting, whenever possible, empyrical
data. This will allow an isolation of the effect of the dust distribution 
on the FIR emission. We will try to explain the characteristic of FIR
emission with diffuse dust heated by a diffuse ISRF. Clumping of dust
and embedded stellar emission \citep{BianchiSub1999} are not yet included in 
this work.

Recent FIR observations suggest large scalelengths for dust emission.
In a sample of seven spirals resolved by the ISOPHOT instrument aboard
the ISO satellite \citep{KesslerA&A1996,LemkeA&A1996}, \citet{AltonA&A1998}
found that the 200 $\mu$m scalelengths are 70\% larger than those in the
B-band. The large ratio may be a result of a dust disk larger than the
stellar distribution. However, the result may also be affected by the
transient effects of the ISOPHOT detectors. In fact, the P32 mapping
mode with which the \citet{AltonA&A1998} images were obtained is not yet
scientifically validated \citep{KlaasRep2000}. We will use the
model presented here to test if indeed the observations are compatible
with large disks. Other works seems to support the same hypothesis. 
\citet{DaviesMNRAS1997} could model the Galactic FIR surface brightness
only with a dust disk with radial and vertical scalelengths larger than 
the stellar by a factor 1.5 and 2.0, respectively.
\citet*{NelsonAJ1998} combined 100 $\mu$m IRAS images of galaxies
with similar angular size and concluded that dust is present well beyond
the observed stellar disk. \citet{XilourisSub1998} derived
intrinsic dust scalelengths larger than the stellar one, by a mean
factor 1.4 in the seven edge-on spiral sample. If confirmed, such 
disks may have a large impact on the observations of the distant
universe, because of their larger cross section to radiation from
backgroud objects.

The paper is structured as follows: Sect.~\ref{program} describes the
radiative transfer code, with a particular attention to the chosen
dust and stellar distributions and to the modification with respect to
the original BFG code. Sect.~\ref{mirsect} and \ref{fircode} show 
how the map of absorbed energy derived from the radiative transfer model
can be converted into a map of the dust temperature distribution and hence,
of FIR emission. We then show an application of the model to the spiral
galaxy NGC~6946 with the intent of analysing the behaviour of our model
and test the influence of the various parameters on the FIR emission.
The observed properties of NGC~6946 used in the model are described
in Sect.~\ref{model}. The results are presented in Sect.~\ref{results} and 
discussed in Sect.~\ref{discussion}. A summary is given in the last section.

\section{The radiative transfer model}
\label{program}

The original BFG radiative transfer code carries out the radiative transfer 
for typical galactic geometries. Simulation of optical surface brightness,
as well as light polarisation, are produced. Dust properties are computed 
from the \citet{DraineApJ1984} dust model, using Mie's theory for spherical
grains. In this paper we use a simplified version of the code: the
treatment of polarization has been omitted; empirical dust properties 
and scattering phase functions are used.
The same approach has recently been adopted in \citet{FerraraApJS1999} and 
\citet{BianchiSub1999}. 

In this Section, we describe the stellar and dust distributions adopted
in the galaxy model, and the chosen dust extinction and scattering 
properties. A brief description is then given of the radiative transfer 
code.

\subsection{The stellar disk}
\label{stardisk}

As in BFG, the new version of the code allows the use of both disks and 
spheroidal bulges to describe the spatial distribution of stars
\citep{BianchiThesis1999}. However, to limit the number of free parameters 
in the modelling, we adopt here a single stellar disk. The chosen disk
is exponential along both the radial and vertical directions. The 
luminosity density distribution is thus described by 
\begin{equation}
\rho=\rho_0 \mbox{exp}(-r/\alpha_\star-\mid z\mid/\beta_\star).
\label{vertexpo}
\end{equation}
where r and z are the galactocentric distance and the height above the
galactic plane, respectively, and $\alpha_\star$ and $\beta_\star$ the
relative scalelengths. While there is a consensus for the radial 
exponential behaviour~\citep{DeVaucouleursBook1959,FreemanApJ1970},
a number of expressions for the vertical distribution have been used in
the literature.  \citet{VanDerKruitA&A1981} use a sech$^2$, the
solution for a self gravitating isothermal sheet, while
\citet{VanDerKruitA&A1988} propose a sech, consistent with measures of
stellar velocity dispersion out of the Galactic plane. 
The exponential adopted here \citep*{WainscoatApJ1989} has the advantage 
of mathematical simplicity, but has no firm physical justification.
Analysing the vertical profiles of a sample of 24 edge-on galaxies 
in the relatively dust-free K-band, \citet*{DeGrijsA&A1997b} find 
that a distribution with a peak intermediate between the exp and the 
sech fits better the central peak~\citep[see also][]{DeGrijsA&AS1996}. 
Since a small inclination from the pure edge-on case can produce a 
less sharp profile, these results are consistent with an exponential 
distribution. 

The radiative transfer model of the galaxy (Sect.~\ref{montecarlo}) is 
run for several wavelength bands. The radial scalelength of the stellar 
distribution, $\alpha_\star$, is assumed to be the same for all of these
bands. Therefore, any observed colour gradient along the disk of a
spiral is interpreted as due to differential extinction with $\lambda$,
rather than being a reflection of different stellar components. 
The wavelength dependence of the intrinsic stellar radial scalelength is 
uncertain, because a knowledge of the dust distribution is needed for
its determination. \citet{PeletierA&A1995} analyzed the variation of the 
ratio of B and K-band scalelengths with inclination in a sample of 37 Sb-Sc 
galaxies and concluded that the observed gradient is due to extinction. 
On the other hand, \citet{DeJongA&A1996b} compared colour-colour plots of a 
sample of 86 face-on galaxies with a Monte Carlo radiative transfer model 
and finds that {\em reasonable} dust models cannot be responsible for the 
observed colour gradients. However, such a conclusion does not seem to 
include models with dust disks more extended that the stellar, that is 
one of the cases studied here. The fits of edge-on galaxies of 
\citet{XilourisSub1998} show a slight variation of the intrinsic scalelength 
with wavelength.

Observations on the Galaxy show that different stellar populations have 
different vertical scalelengths. For the sake of simplicity, a single mean 
value for the vertical scalelength has been adopted here, for each wavelength.
\citet{WainscoatApJS1992} provide a table of vertical scalelengths, B,
V, J, H, K-band absolute magnitudes and local number density for the
main Galactic stellar sources. A mean value has been computed averaging
over the total disk luminosity integrated from the B to the K band.
Assuming $\alpha_\star=3$ kpc \citep*{KentApJ1991,FuxA&A1994},
we derive $\alpha_\star/\langle\beta_\star\rangle=14.4$. Similar values for 
$\alpha_\star/\langle\beta_\star\rangle$ can be found in the literature 
\citep{VanDerKruitA&A1982,DeGrijsA&AS1996,XilourisSub1998}. 

The stellar disk is truncated at 6$\alpha_\star$. Fits of star counts for
faint Galactic sources suggest a similar truncation
(\citealt{WainscoatApJS1992}; \citealt*{RobinA&A1992};
\citealt{RuphyA&A1996}). The truncation along the vertical direction is
at 6$\beta_\star$.

\subsection{The dust disk}
\label{dustdistri}

The parameters for the dust disk are far more uncertain than those
for the stellar disk. Usually the same functional form is used
as for the stellar distribution, with independent scalelengths 
\citep{KylafisApJ1987,ByunApJ1994,BianchiApJ1996,XilourisSub1998}.

A very good correlation is usually found between dust and gas 
in spirals, especially with the molecular component which is dominant
in the inner galaxy \citep{DevereuxApJ1990,XuA&A1997,BianchiMNRAS1998,
AltonApJL1998,BianchiA&A2000}. It is therefore reasonable to use the 
gas distribution as a tracer of dust.

In luminous, face-on, late-type spirals, H$_2$ peaks in the centre
and falls off monotonically with increasing galactocentric distance,
while the HI distribution shows a central depression and a nearly
constant surface density across the rest of the optical disk
\citep{YoungARA&A1991}. Some early type galaxies show the same behaviour, 
although a good fraction of them presents a central depression and a 
flatter ring distribution for the molecular gas. Indeed, flat or
ring-like dust distributions have been derived from FIR and sub-mm
observations \citep{XuApJ1996b,SodroskiApJ1997,BianchiMNRAS1998}. 

Nevertheless, the gas distribution of most galaxies is dominated by
a centrally peaked molecular gas component. This is the case of 
the late-type spiral NGC 6946 \citep{TacconiApJ1986}, whose dust 
distribution is the main concern of this paper (Sect.~\ref{model}). 
Therefore, in this paper dust is assumed to be distributed in a smooth 
radial and vertical exponential disk, similar to the stellar one 
(Sect.~\ref{stardisk}). The number density of dust grains can be written as 
\begin{equation}
n(r,z)= n_0 \exp (-r/\alpha_d -\mid z\mid/\beta_d),
\label{dustexp}
\end{equation}
with $n_0$ the central number density. The dust scalelengths $\alpha_d$ 
and $\beta_d$ can be selected independently from the analogous stellar 
parameters. It is usually assumed in radiative transfer models that 
$\alpha_d\approx\alpha_\star$ and $\beta_d\approx 0.5\beta_{\star}$ 
\citep[][and references therein]{ByunApJ1994,BianchiApJ1996,DaviesMNRAS1997}.
We will refer to this as the {\em standard model}. The choice of the 
vertical scalelengths is mainly dictated by the impossibility of simulating the 
extinction lanes in edge-on galaxies with a dust distribution thicker than 
the stars \citep{XilourisSub1998}. The assumption of a similar radial
scalelength for stars and dust has no firm justification.
As already said in the introduction, recent analysis of extinction of
starlight and FIR emission suggests more extended dust distributions.

The central number density of dust $n_0$ is normalised from the chosen
optical depth of the dust disk. For a central face-on optical depth in 
the V-band, $\tau_V$, $n_0$ is given by
\begin{equation}
n_0=\frac{\tau_V}{2 \beta_d \sigma_\mathrm{ext}(V)},
\label{prem3}
\end{equation}
where $\sigma_\mathrm{ext}(V)$ is the extinction cross section.  The 
optical depth at any wavelength can then be computed by integrating the 
absorption coefficient
\begin{equation}
k_\lambda (r,z)=n(r,z) \sigma_\mathrm{ext}(\lambda)
\label{abscoeff}
\end{equation}
along any path through the dust distribution. The integral involves knowledge 
of the ratio $\sigma_\mathrm{ext}(\lambda)/\sigma_\mathrm{ext}(V)$, which is 
found from the assumed extinction law. As for the stellar disk, the dust disk 
has been truncated at 6 scalelengths both in the vertical and radial direction. 

In section~\ref{results} various values of $\alpha_d/\alpha_\star$, 
$\beta_d/\beta_\star$ and $\tau_V$ will be explored. 

\subsection{Dust extinction}
\label{assu_ext}

The extinction law adopted here is that typical of diffuse Galactic dust, 
with a strong bump at 2175\AA. Regions of high star formation and 
starburst galaxies usually have a weaker bump and a steeper far-UV rise
\citep*{WhittetBook1992,CalzettiApJ1994,GordonApJ1997}.
Extinction laws in the optical show smaller differences. From radiative 
transfer models of seven edge-on galaxies, \citet{XilourisSub1998} derive 
an extinction law similar to the Galactic one longward of the U-band.

The extinction law used is tabulated in Tab.~(\ref{optpar}), for 17 bands in 
the spectral range of stellar emission. The bands from UV1 to K have 
essentially the same spectral coverage as the homonymous described by 
\citet{GordonApJ1997}, from which the extinction law is taken.
Two bands (namely EUV and LMN) have been added to extend the spectral 
coverage in the ultraviolet up to the ionization limit and in the near 
infrared. The extinction law for these two bands has been taken from 
\citet{WhittetBook1992} and \citet{RiekeApJ1985}, respectively. 

We have used the \citet{HenyeyApJ1941} phase function for the
scattering. This is given by
\begin{equation}
\phi(\theta)=\frac{1}{2}\frac{1-g^2}{(1+g^2-2g\cos\theta)^{3/2}},
\label{hgphase}
\end{equation}
with $\theta$ the angle of scattering. The asymmetry parameter $g$, as 
well as the dust albedo $\omega$, are derived from models of dust scattering 
by reflection nebulae in the Milky Way. The adopted $g$ and $\omega$ are shown 
in Tab.~(\ref{optpar}). Again, the values for the bands from UV1 to K are
from \citet{GordonApJ1997}. For the EUV band we have used
\citet{WittApJ1993} data at 1000\AA\, while the values for the LMN band
have been assumed equal to those in the K band.

\begin{table*}
\centerline{
\begin{tabular}{lcccccc}
\hline
band&\multicolumn{2}{c}{$\lambda_\mathrm{i}<\lambda$ [\AA]$<\lambda_\mathrm{f}$}
& $\sigma_\mathrm{ext}(\lambda)/\sigma_\mathrm{ext}(V)$ & $\omega$ & $g$ & 
$F_\mathrm{MIR}(\lambda)$\\
\hline
EUV & 912 & 1125 & 4.34 & 0.42 & 0.75 & 0.72 \\
UV1 & 1125 & 1375 & 3.11 & 0.60 & 0.75 & 0.54 \\
UV2 & 1375 & 1655 & 2.63 & 0.67 & 0.75 & 0.44\\
UV3 & 1655 & 1900 & 2.50 & 0.65 & 0.73 & 0.47\\
UV4 & 1900 & 2090 & 2.78 & 0.55 & 0.72 & 0.58\\
UV5 & 2090 & 2340 & 3.12 & 0.46 & 0.71 & 0.65\\
UV6 & 2340 & 2620 & 2.35 & 0.56 & 0.70 & 0.54\\
UV7 & 2620 & 3230 & 2.00 & 0.61 & 0.69 & 0.43 \\
U   & 3230 & 3930 & 1.52 & 0.63 & 0.65 & 0.36\\
B   & 3930 & 4979 & 1.32 & 0.61 & 0.63 & 0.28\\
V   & 4979 & 5878 & 1.00 & 0.59 & 0.61 & 0.23\\
R   & 5878 & 7350 & 0.76 & 0.57 & 0.57 & 0.18\\
I   & 7350 & 9500 & 0.48 & 0.55 & 0.53 & 0.14\\
J   & 9500 & 14000  & 0.28 & 0.53 & 0.47 & 0.12\\
H   & 14000& 19000 & 0.167& 0.51 & 0.45 & 0.12\\
K   & 19000& 25000 & 0.095& 0.50 & 0.43 & 0.12\\
LMN & 25000& 37500 & 0.04 & 0.50 & 0.43 & 0.12\\\hline
\end{tabular}
}
\caption{Milky Way extinction law 
($\sigma_\mathrm{ext}(\lambda)/\sigma_\mathrm{ext}(V)$), albedo $\omega$ and 
asymmetry parameter $g$ and wavelength range of the 17 bands used in
the models. The last column gives the value of the MIR correction
$F_\mathrm{MIR}$ (Eqn.~\ref{mircorre}).}
\label{optpar}
\end{table*}

\subsection{The Monte Carlo code}
\label{montecarlo}

We give here a brief description of the radiative transfer code, referring 
the interested reader to the BFG paper for more details.

Adhering to the Monte Carlo method, the code follows the life of each
energy unit (a \emph{photon}) through scattering and absorption, until the 
radiation is able to escape the dusty medium. The code is monochromatic, 
since the optical properties of dust must be specified for a particular 
wavelength. In principle, the geometrical distributions of stars may depend 
on $\lambda$ too. For a given star-dust geometry and central face-on
optical depth $\tau_V$, a radiative transfer simulation is produced for each 
of the 17 bands of Sect.~\ref{assu_ext}. Typically, a simulation consists of
10$^7$ photons.

The main steps of the computation scheme are:
\begin{description}
\item[Emission:] the position of a photon in the 3-D space is derived 
according to the stellar distributions described in Sect.~\ref{stardisk}. 
Photons are emitted isotropically and with unit energy.
\item[Calculation of optical depth:] The absorption coefficient 
$k_\lambda (r,z)$ (Eqn.~\ref{abscoeff}) is integrated from the emission 
position along the photon travelling direction to derive the total optical 
depth $\tau_T$ through the dust distribution.  A fraction $e^{-\tau_T}$ of 
all the energy travelling in that direction propagates through the
dust. With the Monte Carlo method it is then possible to extract the optical 
depth $\tau$ at which the photon impinges on a dust grain. 
This optical depth can be computed inverting
\begin{equation}
\int_0^{\tau} e^{-\xi} d\xi=R,
\end{equation}
where $R$ is a random number in the range [0,1]. If the derived $\tau$
is smaller than $\tau_T$, the photon suffer extinction, otherwise it
escapes the dusty medium. This process is quite inefficient when the
optical depth of the dust distribution is small, most of the photons
leaving the dust distribution unaffected. To overcome this problem, the
{\em forced scattering} method is used \citep{CashwellBook1959,WittApJS1977}:
essentially, a fraction $e^{-\tau_T}$ of the photon energy is unextinguished 
and the remaining $1-e^{-\tau_T}$ is forced to scatter.
When the optical depth is small ($\tau_T<10^{-4}$) or the photon path is
free of dust, the photon escapes the cycle.
Once $\tau$ is known, the integral of $k_\lambda (r,z)$ is inverted to
derive the geometrical position of the interaction between the photon and
the dust grain.
\item[Scattering and Absorption:] a fraction of the photon energy, given
by the albedo $\omega$, is scattered, while the remaining $(1-\omega)$
is absorbed. The scattering polar angle $\theta$, i.e. the angle between
the original photon path and the scattered direction, is computed using
the \citet{HenyeyApJ1941} scattering phase function (Eqn.~\ref{hgphase}),
inverting
\begin{equation}
\int_0^\theta \phi(\theta') \sin\theta'= R,
\label{phasinv}
\end{equation}
with $R$ a random number. The inversion of Eqn.~(\ref{phasinv}) is given
by the analytical formula
\begin{equation}
\theta=\arccos\left[\frac{1}{2g}\left(1+g^2-\frac{(1-g^2)^2}{(1+g(1-2R))^2}
\right)\right].
\end{equation}
No preferential direction perpendicular to the original photon path
is assumed, as for scattering by spherical grains. The azimuthal angle is
thus extracted randomly in the range $[0,2\pi]$.

The amount of energy absorbed by the dust grain is stored as a function of 
the galactocentric distance and the height above the plane, using the two
model symmetries, i.e. the symmetry around the vertical axis and the
symmetry about the galactic plane, to improve the signal-to-noise. 
Maps of absorbed energy were not produced by the original BFG code.
\item[Exit conditions:] the last two steps are then repeated, using
the new direction of the scattered photon, the coordinates of the 
scattering point and the energy reduced by absorption,
until the energy of the photon falls below a
threshold value ($10^{-4}$ of the initial energy) or until the exit 
conditions on $\tau$ are verified.
\end{description}
After the exit conditions are satisfied, the photon is
characterised by the last scattering point, its travelling direction and
its energy. To reduce the computational time, the two model symmetries are 
exploited to produce a total of 4 photons from each one. 
Photons are then classified according to the angle between the last direction 
and the symmetry axis. An image is produced collecting all the photons that 
fall in a solid angle band of width $4\pi/15$ (BFG) and mean polar angle 
equal to the chosen model inclination.

Finally, all the photons in an angle band are projected into the plane of 
the sky according to their point of last scattering. Maps of 201x201
pixels are produced, covering an area of 12x12 stellar radial scalelengths 
around the centre of the galaxy. Maps of absorbed energy cover 6 dust 
scalelengths in the radial direction and in the positive vertical direction 
in 101x101 pixels.

\subsection{Normalisation of the radiative transfer output}
\label{SED}

Each set of 17 radiative transfer simulations is then normalised
according to a chosen SED for the stellar radiation. Two kinds of 
normalization are available: in the first, the \emph{input} 
normalization, a SED is chosen for the intrinsic, unextinguished,
stellar energy emission. The SED is integrated over the bands limits
and the total energy emitted in a band is used to scale each
monochromatic simulation. This is suitable, for instance, to predict 
the FIR emission associated with a synthetic galactic SED.

The second normalisation mode, the \emph{output} normalisation, is
instead based on the observed SED of a galaxy. An observed SED is
constructed from observations in several wavelength and the model is
scaled to produce images that have the same SED. The intrinsic,
unextinguished, SED is then inferred from the radiative transfer model.
This second mode, suitable for fitting the FIR emission, is the one used
in this paper (Sect.~\ref{model}).

Together with the optical images, the absorbed energy maps are also normalised.
Thus, the output of a model consists of a set of 17 images in the 
wavelength range of stellar emission and a set of 17 maps of the energy 
absorbed by dust from stellar radiation emitted in each band. The images are
produced in units of surface brightness, while the maps of absorbed energy are 
in units of energy per unit time per unit volume.
The absorbed energy maps can be added together to produce a map of
the total energy absorbed by dust. However, for the purpose of
modelling the FIR emission, a further step is needed before coadding.
This is explained in the next Section.

\section{The MIR correction}
\label{mirsect}

The energy absorbed from photons heats up dust grains and is re-emitted in 
the infrared, preferentially at $\lambda > 10\mu$m. We do not consider
here the heating of dust grains by collisions with the interstellar gas: 
this process is normally negligible and the dust temperature is almost 
entirely determined by radiative processes \citep{SpitzerBook1978}.

When the energy of the absorbed photon is small compared to the internal 
energy of the grain, radiation is emitted at the thermal equilibrium. This
is the case for larger grains, responsible for most of the FIR emission. For
smaller grains, the absorption of a single high-energy photon can 
substantially alter the internal energy. The grain undergoes temperature
fluctuations of several degrees and cools by emission of radiation,
mainly in the MIR range \citep{WhittetBook1992}. The models of this
paper are limited to thermal FIR emission only. Therefore, it is necessary
to exclude from the total absorbed energy the fraction that goes into
non-equilibrium heating.

\begin{figure}
\resizebox{\hsize}{!}{\includegraphics{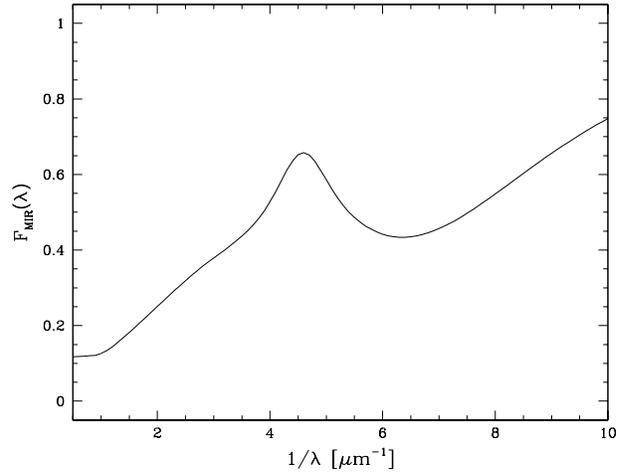}}
\caption{The contribution of very small grains and PAHs to the absorption, 
$F_\mathrm{MIR}(\lambda)$ (Eqn.\ref{mircorre}), derived from the 
\citet{DesertA&A1990} model.}
\label{fig_desert}
\end{figure}

\citet*{DesertA&A1990} derived an empirical dust model, from an analysis 
of the features in the Galactic extinction law and in the infrared emission.
Three dust components were needed in the model: i) big grains 
($0.015\mbox{$\mu$m}<a<0.11\mbox{$\mu$m}$), responsible for the FIR emission; 
ii) very small grains ($0.0012\mbox{$\mu$m}<a<0.015\mbox{$\mu$m}$) and iii) 
PAHs, responsible for the MIR emission at $\lambda<80\mbox{$\mu$m}$. 
From the parameters and functional forms of \citet{DesertA&A1990}, we have 
derived the mean absorption cross-section of the model, 
$\sigma_\mathrm{abs}(\lambda)$, and for each of its three dust 
components\footnote{Note that
$\sigma_\mathrm{ext}=\sigma_\mathrm{abs}+\sigma_\mathrm{sca}$. In the
\citet{DesertA&A1990} model, very small grains and PAHs are pure absorbers 
(i.e. $\sigma_\mathrm{sca}=0$).}.  Of the light impinging on the grain 
mixture, a fraction of energy proportional to $\sigma_\mathrm{abs}(\lambda)$ 
is absorbed; therefore, the contribution of very small grains and PAHs to the 
absorption is given by
\begin{equation}
F_\mathrm{MIR}(\lambda)=\frac{\sigma_\mathrm{abs}^\mathrm{VSG}(\lambda)+
  \sigma_\mathrm{abs}^\mathrm{PAHs}(\lambda)}{\sigma_\mathrm{abs}(\lambda)}.
\label{mircorre}
\end{equation}
Values of $F_\mathrm{MIR}(\lambda)$, the MIR correction, are given in 
Tab.~(\ref{optpar}) for each of the model bands and plotted in 
Fig.~\ref{fig_desert} as a function of $1/\lambda$.
$F_\mathrm{MIR}(\lambda)$ has a local maximum in the position of the
2175\AA\ bump of the extinction curve: absorption by very small grains 
in the \citet{DesertA&A1990} model is responsible for this feature. The
rise in the Far-UV is due to PAHs.

After multiplying by $(1-F_\mathrm{MIR}(\lambda)$), the 17 absorbed energy 
maps store the energy that is absorbed by large grains only.
The final coadded map, $W_\mathrm{abs}(r,z)$, contains the energy absorbed 
by dust per unit time and unit volume, that goes into thermal
equilibrium emission only. 
The derivation of the dust temperature and FIR emission from
$W_\mathrm{abs}(r,z)$ is shown in the next Section.

\section{The dust emission model}
\label{fircode}
The power absorbed by a single dust grain can be derived by dividing
$W_\mathrm{abs}(r,z)$ by the grain number density $n(r,z)$. Because of
the conservation of energy, this is equal to the power each grain
radiates. Assuming that all dust grains in our model are spherical and
with the same mean radius $a$, the thermal equilibrium imposes that 
\begin{equation}
\frac{W_\mathrm{abs}(r,z)}{n(r,z)}= 4\pi a^2 \int_0^\infty
Q_{\mathrm{em}}(\lambda)\pi B_\lambda(T_{\mathrm{d}}(r,z))d\lambda,
\label{prem1}
\end{equation}
where $T_{\mathrm{d}}(r,z)$ is the dust temperature distribution and 
$Q_{\mathrm{em}}(\lambda)$ the emission efficiency. By Kirchhof's law,
$Q_{\mathrm{em}}=Q_{\mathrm{abs}}=\sigma_{\mathrm{abs}}/\pi a^2$
\citep{WhittetBook1992}.

For the exponential dust distribution of Eqn.~(\ref{dustexp}) and the
central number density of Eqn.~(\ref{prem3}), Eqn.~(\ref{prem1}) can be
rewritten as
\begin{eqnarray}
\frac{\beta_d W_\mathrm{abs}(r,z)}{2 \tau_\mathrm{V} \exp (-r/\alpha_d -\mid
z\mid/\beta_d)} =\mbox{\hspace{3.3cm}}\nonumber \\
\mbox{\hspace{3.3cm}}\int_0^\infty \frac{Q_{\mathrm{em}}(\lambda)}{Q_\mathrm{ext}(V)} \pi B_\lambda(T_{\mathrm{d}}(r,z))d\lambda,
\label{prem4}
\end{eqnarray}
with $Q_{\mathrm{ext}}(V)=\sigma_{\mathrm{ext}}(V)/\pi a^2$. The map of dust 
temperature $T_{\mathrm{d}}(r,z)$ is derived by inverting Eqn.~(\ref{prem4}). 

For the ratio $Q_\mathrm{em}(\lambda)/Q_\mathrm{ext}(V)$, we use a
value derived for Galactic dust \citep*{BianchiA&A1999}, 
\begin{equation}
\frac{Q_{\mathrm{em}}(\lambda)}{Q_{\mathrm{ext}}(V)}=\frac{2.005}{2390}
\frac{\left(\mbox{100 $\mu$m}/\lambda\right)^2}{
\left[1+\left(\mbox{200 $\mu$m}/\lambda\right)^6\right]^{1/6}
}.
\label{qema_used}
\end{equation}
Eqn.(\ref{qema_used}) has also been derived assuming that all grains have the 
same radius.  The wavelength dependence of the ratio changes 
smoothly from $\lambda^{-1}$ to $\lambda^{-2}$ at 200 $\mu$m, as observed by
\citet{ReachApJ1995} using high S/N FIR spectra of the Galactic plane.
The absolute value of the ratio has been derived using dust column 
density maps calibrated to Galactic extinction \citep*{SchlegelApJ1998}. 

Finally, FIR images are created integrating the emission coefficient,
\begin{eqnarray}
j_\lambda(r,z)=n(r,z) \sigma_\mathrm{abs}(\lambda) 
B_\lambda(T_{\mathrm{d}}(r,z)) \mbox{\hspace{2.7cm}} \nonumber \\ 
=\frac{\tau_\mathrm{V}}{2\beta_\mathrm{d}} 
\exp (-r/\alpha_d -\mid z\mid/\beta_d)
\frac{Q_{\mathrm{em}}(\lambda)}{Q_\mathrm{ext}(V)}
B_\lambda(T_{\mathrm{d}}(r,z)),
\label{inteff}
\end{eqnarray}
along a given line of sight through the dust distribution, under the 
assumption that dust is optically thin to FIR radiation. Using the 
emissivity of Eqn.~(\ref{qema_used}) this is justified for any model 
with reasonable values of $\tau_\mathrm{V}$.

As for the optical images, the far infrared images are produced in units 
of surface brightness. FIR images have the same extent and resolution of the 
optical images, i.e. a region of 12x12 stellar radial scalelengths around 
the centre of the galaxy is mapped in 201x201 pixels.

\section{A worked example: NGC~6946}
\label{model}

In the following we show an application of the radiative transfer and dust 
emission model to the spiral galaxy NGC~6946. The final goal is to find
the parameters of the dust distribution compatible with the available 
observations in the optical and FIR range.

\begin{table*}
\centerline{
\begin{tabular}{lcl}
 RA (J2000)  & 20$^\mathrm{h}$ 34$^\mathrm{m}$ 52$\fs$0 & \citet{RC3}, RC3 \\
 Dec.        &+60$\degr$ 9$\arcmin$ 15$\arcsec$         &   RC3      \\
 inclination &  34$\degr$    &\protect{\citet{GarciaGomezA&AS1991}}\\
 P.A.        & 64$\degr$    &\protect{\citet{GarciaGomezA&AS1991}}\\
 distance    &  5.5 Mpc & \citet{TullyBook1988} \\
 D$_{25}$    &  11.5$\arcmin$ &  RC3 \\
 $A_B$       & 1.73           &  RC3 \\
 L$_\mathrm{dust}$& 1.5 - 3 $\cdot 10^{10}$ L$_\odot$ & this work,
 Sect.~\ref{results}\\
 M$_\mathrm{gas}$ & $9\cdot 10^9$ M$_\odot$ &
 \citet{DevereuxApJ1990}, rescaled to 5.5Mpc
\end{tabular}
}
\caption{Basic properties of NGC 6946.}
\label{tab_basic}
\end{table*}

NGC~6946 is a large nearby Sc galaxy (A few basic properties are presented 
in Table~\ref{tab_basic}). 
The optical appearance of the galaxy is characterised by six prominent
spiral arms of which the three arms originating from the NE quadrant are 
brighter and more developed than those in the SW \citep{TacconiApJ1990}.
\citet{TrewhellaMNRAS1998,TrewhellaThesis1998} finds that the NE interarm 
region suffers high extinction. The region appear fainter because of the
dust effect, rather than being intrinsically less luminous. Indeed, polarised 
light is observed in the interarm regions as well as in the spiral arms 
\citep{FendtA&A1998}. 

NGC~6946 shows a centrally peaked molecular gas distribution
\citep{TacconiApJS1989} dominant in the inner 10$\arcmin$
over the flatter atomic
gas component \citep{TacconiApJ1986}. The galaxy is marginally resolved in 
FIR observations with KAO and IRAS \citep{EngargiolaApJS1991,AltonA&A1998}, 
with the dust emission tending to follow the spiral arms and the bright 
central emission. The 117$\arcsec$ resolution 200$\mu$m ISO image 
shows a morphology similar to the 100$\mu$m IRAS observation 
\citep{AltonA&A1998}. Higher resolution SCUBA images show a tight
correlation between the dust emission at 850$\mu$m and the dominant gas 
phase \citep{BianchiA&A2000}. 

We now describe the observed SED for the stellar emission, used in the
normalization of the radiative transfer, as shown in Sect.~\ref{SED},
and the SED of dust emission, that will be compared to the model results.
The procedure adopted in the modelling are presented in Sect.~\ref{wopro}.

\subsection{The stellar Spectral Energy Distribution}
\label{starsed}
We have derived the observed stellar SED from literature data.
Values for the flux inside an aperture of 5$\arcmin$
(corresponding to the B-band 
half light radius; \citealt{EngargiolaApJS1991}) are presented in 
Table~\ref{tab_sed}. All the data have been corrected for Galactic 
extinction (Table~\ref{tab_basic}) using the assumed extinction law.

Optical and Near Infrared data in the bands U, B, g, V, r, I, J, H, K are
from \citet{EngargiolaApJS1991}. The observed NIR emission is considered as
purely stellar. Using the IRAS 12 $\mu$m flux as a template of small grain 
emission and the \citet{DesertA&A1990} model, we derived a dust 
contribution to the K-band emission of only 0.3\%. The stellar emission
at 5 $\mu$m has been extrapolated from the K-band, using the
Rayleigh-Jeans spectral region in the synthetic galactic SEDs of
\citet{FiocA&A1997}.

\begin{table}
\centerline{
\begin{tabular}{ccc}
\multicolumn{3}{l}{stellar emission}\\ \hline
 & $\lambda$ &f$_\nu$ (5$\arcmin \diameter$)\\ &$\mu$m& Jy\\ \hline
                & 0.091 & 0.05                \\
\emph{short}-UV & 0.165 & 0.09 $\pm$ 0.02     \\
\emph{medium}-UV& 0.250 & 0.13 $\pm$ 0.09     \\
\emph{long}-UV  & 0.315 & 0.3  $\pm$ 0.2      \\
U               & 0.360 & 0.50 $\pm$ 0.05     \\
B               & 0.435 & 0.90 $\pm$ 0.06     \\
g               & 0.495 & 1.1  $\pm$ 0.1      \\
V               & 0.554 & 1.38 $\pm$ 0.07     \\
r               & 0.655 & 2.0  $\pm$ 0.2      \\
I               & 0.850 & 3.2  $\pm$ 0.3      \\
J               & 1.250 & 5.1  $\pm$ 0.4      \\
H               & 1.650 & 7.2  $\pm$ 0.6      \\
K               & 2.200 & 5.8  $\pm$ 0.5      \\
                & 5.000 & 1.3                 \\
\\
\multicolumn{3}{l}{dust emission}\\ \hline
                & 12    & 11 $\pm$ 2     \\
                & 25    & 14 $\pm$ 3     \\
                & 60    & 120 $\pm$ 20     \\
                &100    & 240 $\pm$ 50     \\
                &160    & 310 $\pm$ 60     \\
                &200    & 280 $\pm$ 80     \\
                &450    & 9 -- 27  \\
                &850    & 2 -- 3   \\
		\hline
\end{tabular}
}
\caption{NGC 6946 mean flux inside a circular aperture of
diameter 5$\arcmin$ (corresponding to the B-band half light radius; 
\citealt{EngargiolaApJS1991}), for several wavelengths.
The origin of the data is described in Sect.~\ref{starsed} and
\ref{dustsed}, for the stellar and dust emission, respectively.}
\label{tab_sed}
\end{table}

The SED in the non-ionising UV is taken from \citet*{RifattoA&AS1995b}.
The authors derive fluxes for a large sample of galaxies, by
homogenising observation from several satellites (notably IUE),
balloon and rocket-borne experiments, with different apertures and
sensitivities. An aperture correction depending on the morphologycal
type is applied to each flux, and total magnitudes  are given for three
photometric bands centred at 1650\AA\ (\emph{short}-UV), 2500\AA\
(\emph{medium}-UV), 3150\AA\ (\emph{long}-UV). Using an appropriate
aperture correction and calibrating as in \citet*{RifattoA&AS1995a}, we 
have derived fluxes for our selected aperture.
The quoted errors come mainly from the aperture correction.
The flux at the Lyman limit (912\AA) has been extrapolated after
observing that in a $\nu f_\nu$ versus $\log \lambda$ plot
the observed SED is flat for the \emph{short}- and \emph{medium}-UV.
This trend is assumed to be valid down to the ionization limit. 
The ionising UV is not included in the model (Sect.~\ref{discussion}).

\subsection{The dust Spectral Energy Distribution}
\label{dustsed}
The SED of dust emission is also shown in Table~\ref{tab_sed}.
Surface brightnesses over the half-light aperture shortward of 
100$\mu$m have been derived from IRAS High Resolution (HiRes) 
images \citep{RiceAJ1993,AltonMNRAS1998}. Derived values have been 
colour corrected \citep{RiceApJS1988} and are consistent (within a 
20\% error; \citealt{AltonA&A1998}) with the analogous data provided by 
\citet{EngargiolaApJS1991}, derived on previous enhanced resolution 
IRAS images. The value at 160$\mu$m has been derived by
\citet{EngargiolaApJS1991} from the air-borne KAO telescope.

Data at 200$\mu$m are derived from P32 images taken from the ISOPHOT 
instrument aboard the ISO satellite \citep{AltonA&A1998}. The P32
mapping mode is still not scientifically validated and its photometric
calibration is highly uncertain \citep{KlaasRep2000}. \citet{AltonA&A1998}
compared both the integrated flux with the value derived by
\citet{EngargiolaApJS1991} on 200$\mu$m KAO images and the measured
background with an extrapolation from the 100$\mu$m value. They 
conclude that the calibration may overestimate the flux by about 30\%.
A 30\% error is shown in Table~\ref{tab_sed}.

Finally, lower and upper limit of the flux at 450 and 850
$\mu$m are derived from SCUBA images \citep{BianchiA&A2000}. The lower
limit correspond to the signal coming from regions 3$\sigma$ brighter
than the sky, inside the selected aperture. The upper limit is derived
assuming a 1$\sigma$ emission for the regions without detected signal.
Sub-mm fluxes are only given for completeness, the FIR models being 
constrained mainly by the fluxes at 100, 160 and 200 $\mu$m. Emission 
at 12, 25 and 60 $\mu$m, dominated by small grains \citep{DesertA&A1990},
is used in Sect.~\ref{discussion} to validate the MIR correction.

\subsection{Modelling procedure}
\label{wopro}

After choosing the star-dust geometry, a radiative transfer simulation is
run for each of 17 bands of the model (Sect.~\ref{SED}) and images for the 
inclination of NGC 6946 are produced (Table~\ref{tab_basic}).
An intrinsic scalelength $\alpha_\star=2.5$ 
kpc has been used for the stellar emission, derived from a K-band image 
\citep{TrewhellaThesis1998}, where the effects of extinction are smaller.
For the $\alpha_\star/\langle\beta_\star\rangle$ ratio of
Sect.~\ref{stardisk}, this leads to $\beta_\star=$170 pc.
The simulations do not depend on the absolute values of the 
scalelengths, as long as they are shown as functions of scaled
galactocentric and vertical distances.  However, absolute values are 
needed if we want to derive correct values for the emitted and absorbed 
energies in each band.

The simulated images are then integrated inside the half light radius 
aperture (derived from the B-band simulation). Mean fluxes over each band 
of Table~\ref{optpar} are obtained by integrating a continuous SED derived 
from the data in Table~\ref{tab_sed}. These fluxes are then used to
scale the results of the aperture photometry on the simulation.
Because of this normalisation, the stellar SED derived from the simulated
images is the same for each model.  The intrinsic \emph{unextinguished} 
energy emitted by stars is then derived from the Monte Carlo code. 
The half-light radius aperture was chosen in the earlier stages of our
analysis because of the availability of one more UV data point in 
\citet{EngargiolaApJS1991}. However, if the  distribution we have adopted is 
a correct description of the galaxy luminosity density, the normalization 
is independent of the aperture dimension, being equivalent to assigning
a value for $\rho_0$ in Eqn.~(\ref{vertexpo}).  After the normalization, FIR 
images for the selected inclination are produced and the fluxes derived, 
as for the images of stellar emission. The simulated SED in the FIR is then 
compared to the observed one. 

For a sample of seven spirals, \citet{AltonA&A1998} measured 
B-band and FIR scalelengths in a galactocentric distance range from 
1.5$\arcmin$ to 3.5$\arcmin$ after smoothing the optical and the IRAS images 
to the poorest resolution of the 200$\mu$m ISO map (FWHM=117$\arcsec$). For 
NGC~6946, they found that the B band scalelength is slightly smaller
than the 200$\mu\mbox{m}$ scalelength, by a factor 0.9, while it is larger
than the 100$\mu$m scalelength, by a factor 1.8. The absolute values of
the scalelengths are presented in Table~\ref{tab_scale}. The large
200$\mu\mbox{m}$ scalelengths, compared to the optical and IRAS data, are a 
general property of the sample. The FIR simulation of this work are
compared with the observations of \citet{AltonA&A1998}. FIR images are
smoothed to the ISO resolution and scalelengths are derived on the same
distance range as the observations\footnote{As described in 
Sect.~\ref{montecarlo}, a region of 12$\alpha_\star$ is covered by 201 
pixels, thus giving a pixel size of 5.7$\arcsec$ (150 pc), for the adopted 
$\alpha_\star$=95$\arcsec$. Therefore the ISO beam can be modelled by a 
gaussian of FWHM$\approx$20 pixels.}. 

Keeping fixed the stellar distribution, several models are produced with 
different parameters for the dust distribution. The final goal is to
obtain a single model able to describe {\em both} the observed FIR SED and 
the spatial distribution of the emission.

\section{Results}
\label{results}

\begin{figure}
\resizebox{\hsize}{!}{\includegraphics[18,158][592,718]{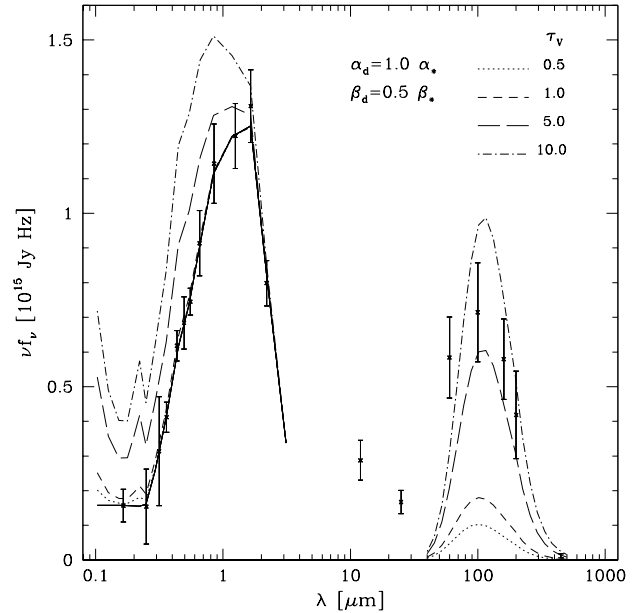}}
\caption{Flux inside the B-band half-light radius for
models with $\alpha_{\mathrm{d}}=\alpha_\star$, $\beta_{\mathrm{d}}=0.5
\beta_\star$ and optical depths $\tau_{\mathrm{V}}=$0.5, 1, 5 and 10.
The data points are those described in Sect.~\ref{starsed} and
\ref{dustsed}, for the stellar and dust emission, respectively.
The solid line represents the SED of stellar emission measured
on the simulated images. For each model,
the lines in the UV-optical-NIR range represent the fluxes
measured for a transparent model with the NGC 6946 inclination.
Because of the quantities plotted, $\nu f_\nu$ vs $\log 
(\lambda)$, the area under the curve is proportional to the emitted
energy.}
\label{sed_standard}
\end{figure}

The first models presented here have the geometrical parameters of the 
{\em standard model}, i.e. a star-dust geometry with
$\alpha_{\mathrm{d}}=\alpha_\star$ and $\beta_{\mathrm{d}}= 0.5
\beta_\star$. As already said in Sect.~\ref{dustdistri}, the choice is mainly 
motivated by the presence of extinction lanes along the major axis of edge-on 
galaxies. The H$_2$ column density in NGC~6946 has an exponential radial 
profile, with a scalelength of 90$\arcsec$ \citep{TacconiApJ1986}.  
Incidentally, 
this is very close to $\alpha_\star$. Thus, the dust disk discussed here can 
be thought of as a dust component associated with the molecular phase.
Four values for the V-band face-on optical depth have been chosen, 
$\tau_{\mathrm{V}}=$0.5, 1, 5 and 10.

The SED of the four models in presented in Fig.~\ref{sed_standard}.
The thick solid line represents the stellar emission of the galaxy as 
derived from the simulations, normalised to the observed data points as
described in the previous section. Each different line shows the intrinsic
unextinguished SED in the short-wavelength part of the spectrum and the
FIR emission in the long-wavelength side. For the dust-free emission,
the SED is derived from the total intrinsic energy, assuming isotropic
emission and measuring the half light radius in a transparent model.
All the models presented in this paper show a spike at $\lambda\approx$2000\AA\
in the unextinguished SED. This is due to the extinction feature at 2175\AA\ 
present in the assumed extinction law (Sect.~\ref{assu_ext}), while the
observed SED is flat. A possible dip in the stellar SED caused by this 
extinction feature could be masked by the broad band of the observation or 
the large errors in the UV photometry. The extinction feature could also 
be absent, in which case our assumption of Galactic dust for NGC~6946 is 
not correct.

In all the models, the wavebands from EUV to U contributes tho 30\% 
of the total absorbed energy, while the wavebands from B to LMN to
60\%. Therefore, most of the radiation is absorbed from the 
Optical-NIR light. A similar result was obtained for the same galaxy
by \citet{TrewhellaThesis1998}.

\begin{figure*}
\begin{minipage}[b]{12cm}
\resizebox{\hsize}{!}{\includegraphics[18,205][592,380]{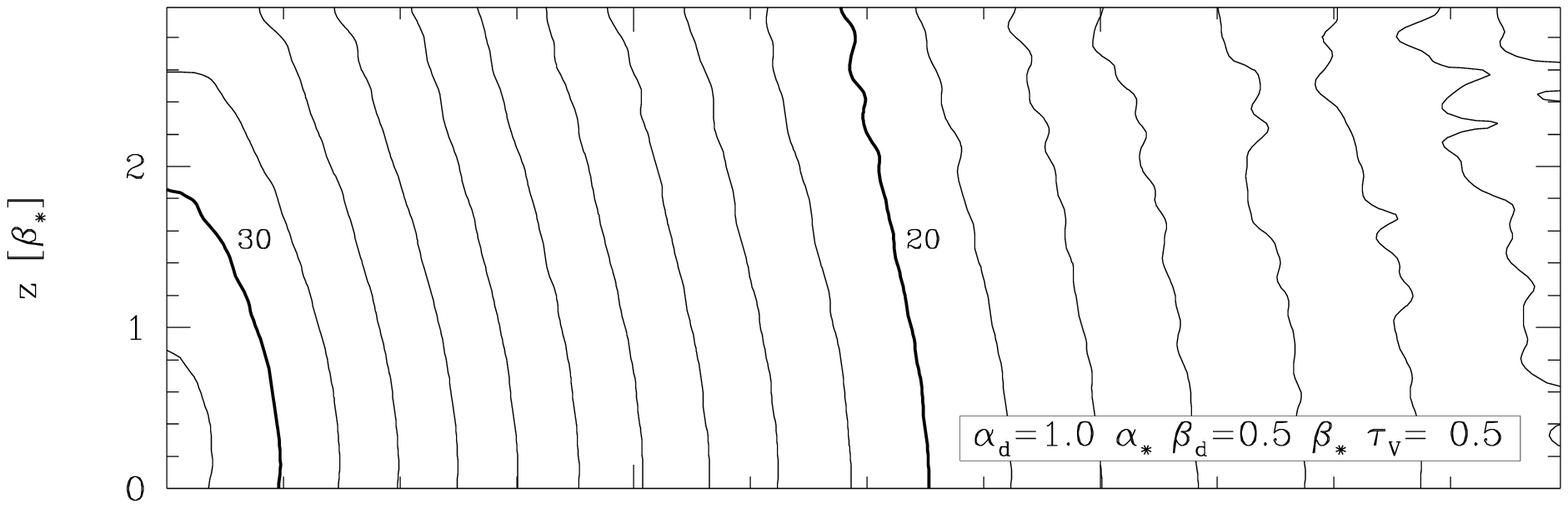}}
\resizebox{\hsize}{!}{\includegraphics[18,205][592,380]{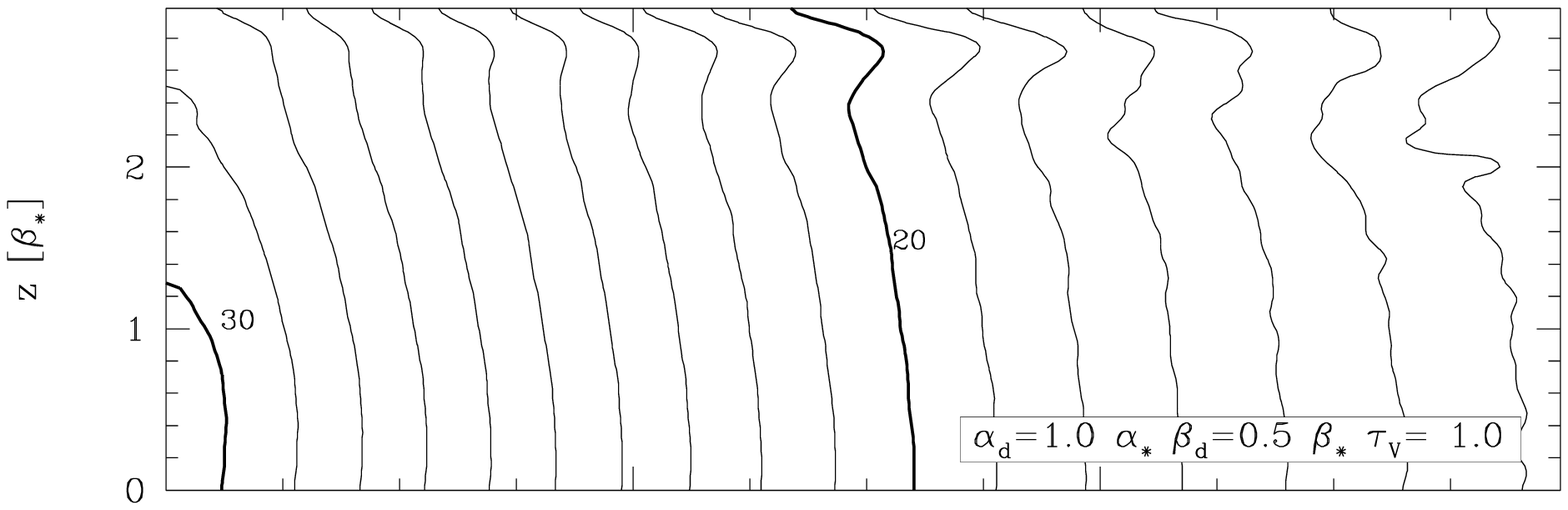}}
\resizebox{\hsize}{!}{\includegraphics[18,205][592,380]{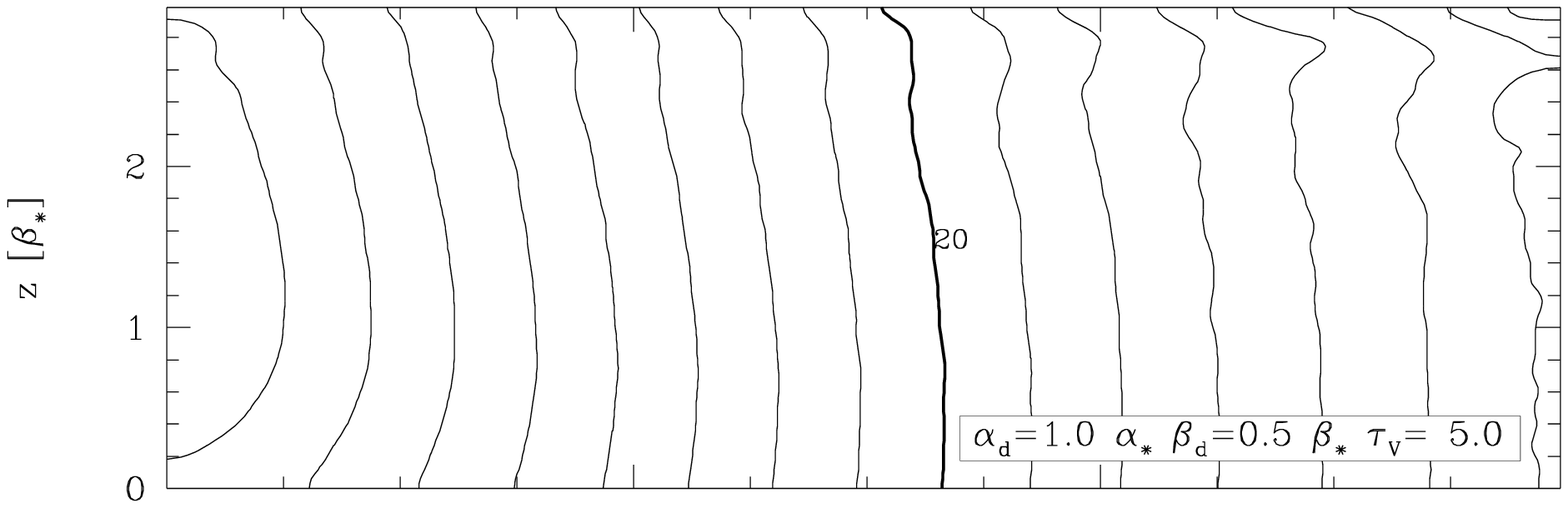}}
\resizebox{\hsize}{!}{\includegraphics[18,168][592,380]{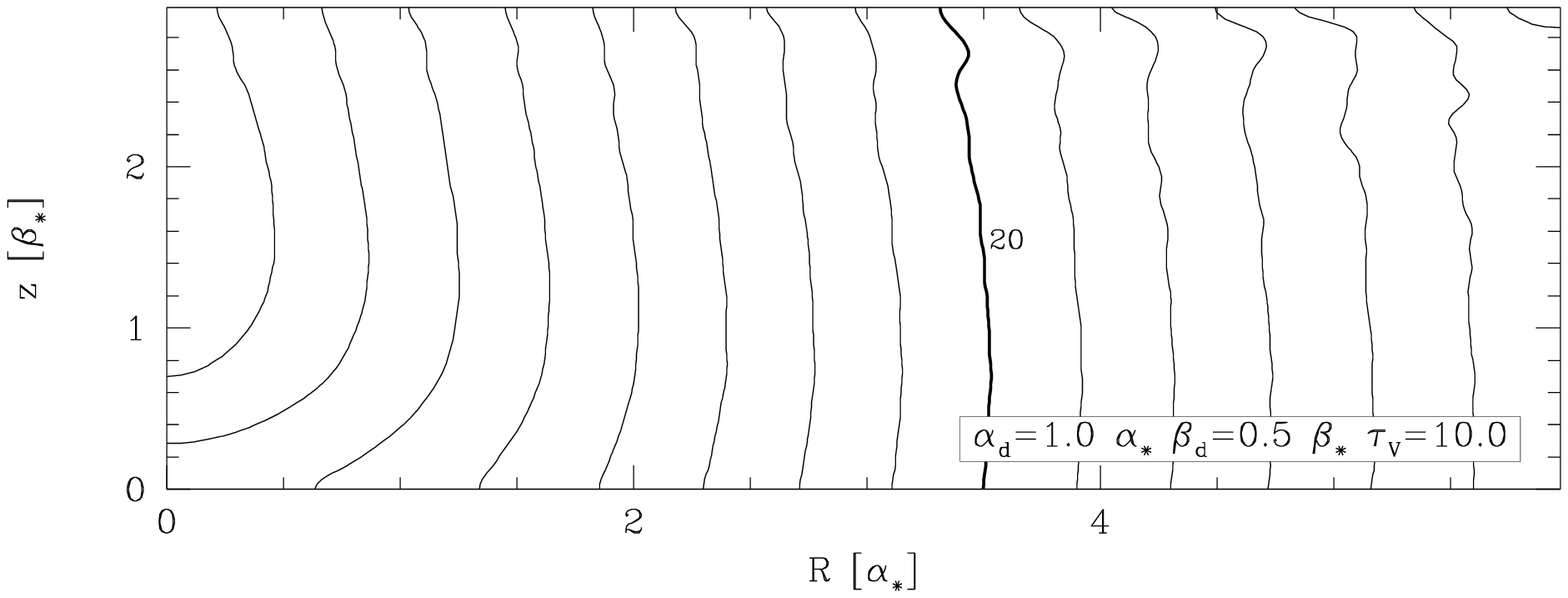}}
\end{minipage}
\hfill
\parbox[b]{55mm}{
\caption{Dust temperature map on a meridian plane for models with
$\alpha_{\mathrm{d}}=\alpha_\star$, $\beta_{\mathrm{d}}=0.5 \beta_\star$ and
optical depths $\tau_{\mathrm{V}}=$0.5, 1, 5 and 10, from top to bottom.
Temperature contours are plotted every 1K and highlighted at 20 K and
30K by a label and a thicker line. The scale along the z-axis has been 
expanded for clarity.}
\label{fig_temp_sta}
}
\end{figure*}

The MIR corrections are quite similar for all the models: approximately
32\% of the total absorbed energy is estimated to go into MIR emission,
the remaining 68\% being available for thermal equilibrium processes
and FIR emission. Since small-grains and PAHs responsible for
no-equilibrium processes have a higher absorption efficiency at
shorter wavelength, the contribution of absorption from Optical-NIR
wavebands is higher after the MIR correction. So the
radiation originally emitted at $\lambda>4000$\AA\ contributes 
$\sim$ 70\% of the FIR emission. The same result applies to all the models
presented in this paper, therefore MIR corrections are not discussed
separately in each case. A comparison between the estimated and observed
MIR emission is given in Sect.~\ref{discussion}.

The temperature distributions for each model are shown in 
Fig.~\ref{fig_temp_sta}, as a function of the galactocentric radius and 
height above the plane. Apart from the central region, the distributions 
are very similar. 
For a dust distribution narrower than the stellar one, the stellar
radiation field is expected to increase with height above the plane
in an optically thick model \citep{DraineApJ1984,RowanRobinsonMNRAS1986}, 
because the stars closer to the plane are shielded. This is evident
in the central regions ($R<1.5\alpha_\star$) of the models. When the
optical depth increases, dust at higher temperature is found at higher
positions above the plane. In the models with higher extinction, the
effect can still be seen at larger galactocentric distances, the region
of higher temperature approaching the galactic plane at large distances. 
Vertical gradients are very shallow, because of the greater extent of
the galaxy in the radial direction with respect to the vertical and because 
the stellar distribution is smooth. 

The FIR spectrum is shown in Fig.~\ref{sed_standard}. Only models with 
central optical depth between $\tau_{\mathrm{V}}$=5 and 10 produce enough 
energy to match the observational data. This is a general property of all the 
models we are going to discuss: a substantial extinction is necessary to
produce the observed SED in the FIR. The total amount of energy
absorbed  and re-emitted (in both MIR and FIR) is between 1.5 and 
$3\cdot 10^{10}$ L$_ \odot$, for the two high $\tau_{\mathrm{V}}$ models,
respectively. This corresponds to a fraction 0.3-0.4 of the intrinsic 
energy produced by the stars. Therefore, $\sim 1/3 $ of the
bolometric luminosity of NGC~6946 is absorbed by dust. 

\begin{table}
\centerline{
\begin{tabular}{rccc}
&$\alpha_\mathrm{B}$& $\alpha_{100\mu\mathrm{m}}$ & $\alpha_{200\mu\mathrm{m}}$ \\ \hline
observed \protect{\citep{AltonA&A1998}} 
		      & 1.52 & 0.86 & 1.66  \\ \\
Standard model $\tau_\mathrm{V}$=0.5 
		      & 1.15 & 0.65 & 0.80 \\
$\tau_\mathrm{V}$=1.0 & 1.19 & 0.66 & 0.81 \\ 
$\tau_\mathrm{V}$=5.0 & 1.63 & 0.72 & 0.86 \\ 
$\tau_\mathrm{V}$=10. & 2.07 & 0.77 & 0.91 \\ \\
$\alpha_\mathrm{d}=1.5\alpha_\star$ model $\tau_\mathrm{V}$=0.5 
		      & 1.16 & 0.76 & 1.01 \\
$\tau_\mathrm{V}$=1.0 & 1.21 & 0.76 & 1.02 \\
$\tau_\mathrm{V}$=5.0 & 1.60 & 0.83 & 1.08 \\ 
$\tau_\mathrm{V}$=10. & 1.76 & 0.87 & 1.12 \\ \\
$\tau_\mathrm{V}$=0.5 HI +$\tau_\mathrm{V}$=5.0 H$_2$ distr.
                      & 1.59 & 0.81 & 1.06 \\ \hline
\end{tabular}
}
\caption{B-band, 100 $\mu$m  and 200 $\mu$m scalelengths for several models, 
in units of $\alpha_\star$ (95$\arcsec$). Scalelengths have been measured 
on the models as described in Sect.~\ref{wopro}. The unsmoothed B-band 
scalelength, measured on a B band image \citep{TrewhellaThesis1998}, is 
1.3$\alpha_\star$.}
\label{tab_scale}
\end{table}

The optical (B) and FIR (100$\mu$m  and 200 $\mu$m) scalelengths measured on 
standard models are shown in Table~\ref{tab_scale}. The B-band scalelength 
increases with $\tau_{\mathrm{V}}$ and it is close to that observed if
$\tau_\mathrm{V}\approx 5$. The FIR scalelengths increase with the
wavelength. A slight increase with $\tau_{\mathrm{V}}$ is also observed, 
because of the smaller temperature in the centre. However, FIR scalelengths
are never as large as observed, especially at 200$\mu$m. While the
100$\mu$m scalelength for the $\tau_\mathrm{V}=5$ model is underestimated by 
$\approx$15\%, the 200$\mu$m scalelength is about one half of that measured 
on ISO images. Therefore, the ratio between the B-band and 200$\mu$m 
scalelengths is smaller than that measured by \citet{AltonA&A1998}.
It is interesting to note that, because of the more rapid increase of the 
B-band scalelength with $\tau_\mathrm{V}$, the lowest values of the ratio 
are obtained for optically thin cases. Nevertheless, optically thin cases 
are unable to explain the large 200$\mu$m scalelength and the energy output.
We note here that, by adopting a single $\alpha_\star$ measured in the 
K-band, we have assumed that the larger optical scalelength with respect to 
that measured in the NIR is entirely due to dust extinction. If indeed an 
intrinsic difference exist, a smaller optical depth may be able to reproduce 
the observed B band scalelength. We will discuss the effect of our assumption 
in Sect.~\ref{discussion}. 

\begin{figure}
\resizebox{\hsize}{!}{\includegraphics[18,158][592,718]{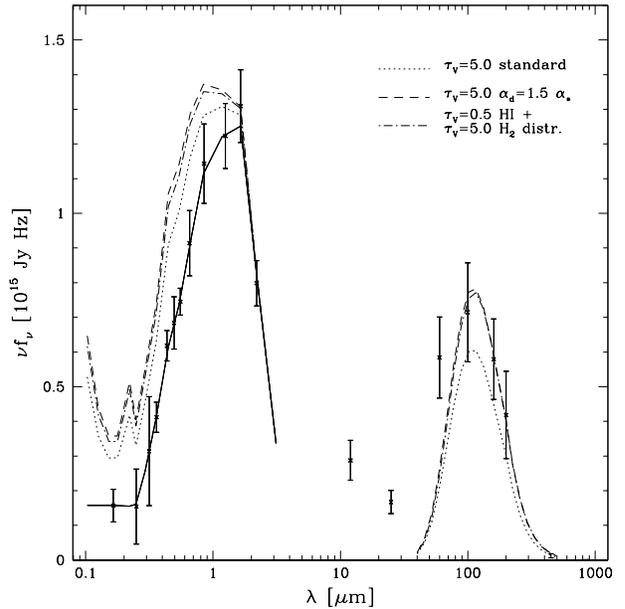}}
\caption{Same as Fig.~\ref{sed_standard}, but for the models with
extended dust distributions. The SED for a standard model with
the same optical depth is also included, for comparison.}
\label{sed_ext}
\end{figure}

\citet{AltonA&A1998} suggest that the large observed scalelengths at 
200$\mu$m is due to a dust distribution more extended than the stellar disk. 
To test this hypothesis, we run models with $\alpha_{\mathrm{d}}= 
1.5 \alpha_\star$ \citep{DaviesMNRAS1997,XilourisSub1998}, keeping the other 
parameters as for the standard model. The $\tau_\mathrm{V}=5$ case is shown 
in Fig.~\ref{sed_ext}. As for standard models, only optically thick cases are 
able to match the observed SED. For the same optical depth the extended model 
has a higher extinction (e.g. 44\% of the energy is absorbed in the V-band 
against the 34\% of the standard model). 

\begin{figure*}
\resizebox{\hsize}{!}{\includegraphics[18,205][592,305]{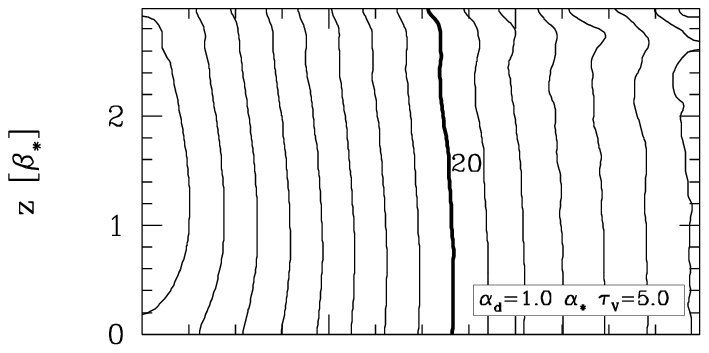}}
\resizebox{\hsize}{!}{\includegraphics[18,205][592,305]{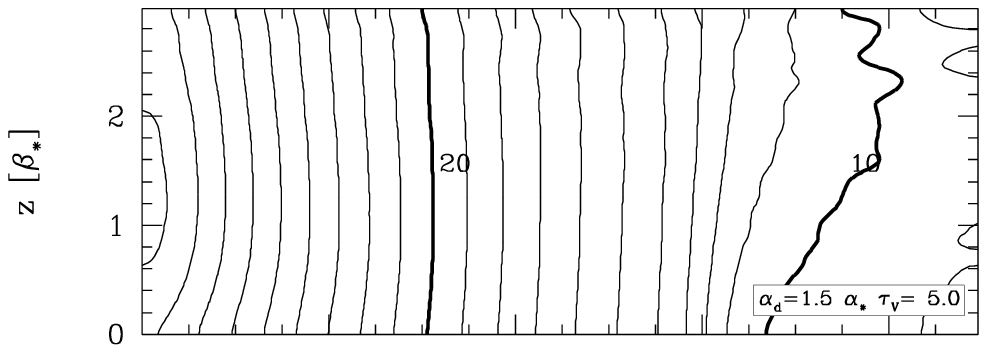}}
\resizebox{\hsize}{!}{\includegraphics[18,170][592,305]{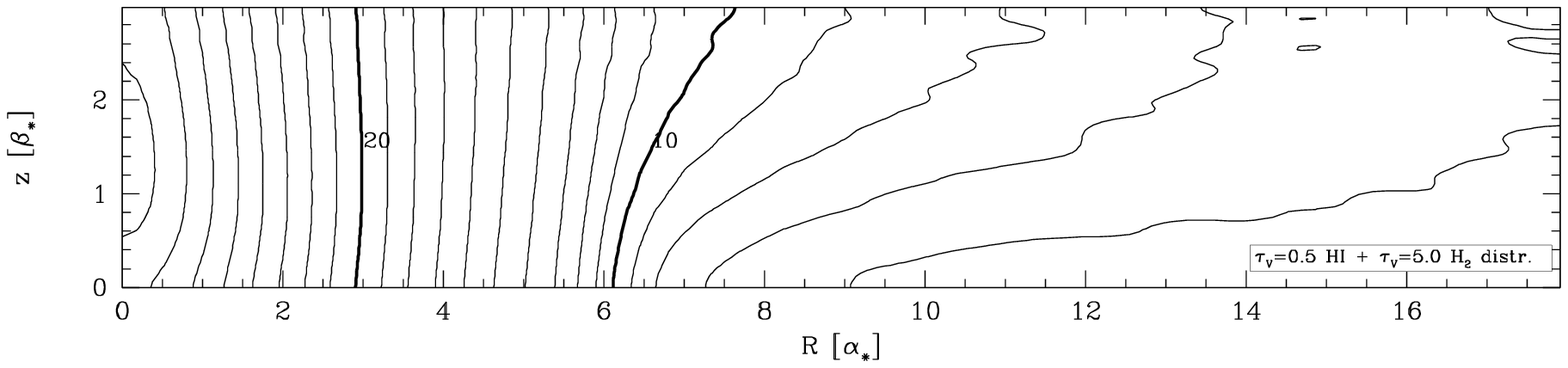}}
\caption{Same as Fig.~\ref{fig_temp_sta}, but for the extended models with
$\alpha_{\mathrm{d}}=1.5\alpha_\star$ and $\tau_{\mathrm{V}}=$5 (central 
panel) and with two dust disk associated to each gas component (lower panel; 
see text for details). The temperature distribution for a standard model with 
$\tau_{\mathrm{V}}=$5 is shown in the top panel, for
comparison. All the models have the same scale along the radial axis.}
\label{fig_temp_ext}
\end{figure*}

The temperature distribution for the extended model is shown in 
the central panel of Fig.~\ref{fig_temp_ext}. For ease of comparison, the 
temperature distribution of the standard model is shown again (in the
right panel), with the same scale as for the new model. Within a radius of 
6 $\alpha_\star$ (the extent of the stellar disk) the temperature pattern of 
the extended model is quite similar, apart from a small difference due to the 
normalisation. This is reflected in the peak of the FIR SED, that is 
essentially the same in both the models. Outside of 6$\alpha_\star$,
the truncation of the stellar distribution, dust is colder and it does not 
modify the shape of the SED.
The steeper temperature gradient at 6$\alpha_\star$ is an artifact due
to the truncation of the stellar disk.  A truncation is indeed suggested by 
counts of faint stellar sources in the Galaxy (Sect.~\ref{stardisk}). A few 
tests have been conducted with stellar distributions truncated at the same 
distance as the dust disks, to avoid having dust in regions without local 
stellar emission. The steeper gradient disappears and a larger distance is 
needed to reach the same temperature.  However, changes are small, the general 
trend in the temperature distribution and in the FIR emission distribution
are essentially the same.
It is interesting to note that for $R>6\alpha_\star$, dust is colder
on the plane than above, because starlight is seen through higher optical
depths along the plane. 

The extended dust distribution causes an increase of the FIR scalelengths with 
respect to the standard model (Table~\ref{tab_scale}). Scalelengths at 
100$\mu$m are quite close to those observed (within a 3\% for the
$\tau_\mathrm{V}=5$ case). However, the 200$\mu$m scalelengths are still 
underestimated, by at least 30\%. B band scalelengths are similar to
those for the standard model (the effect of an extended dust
distribution being appreciable only for high optical depths). 
As in the standard model, despite the increase of the FIR scalelength
with $\tau_\mathrm{V}$, B/200$\mu$m scalelength ratios are close to
those of \citet{AltonA&A1998} only in the optically thin case.

The analysis of  Galactic FIR emission of \citet{DaviesMNRAS1997}
suggests that the vertical scalelength of dust as well is larger than
the stellar one, by a factor of two. Models that included a thicker
dust disk, however, do not produce better results than those presented
here \citep{BianchiThesis1999}.

Larger FIR scalelengths can be produced by extending further the dust
distribution. \citet{DaviesMNRAS1999} note that the IRAS emission tends 
to follow the molecular gas, while the $200\mu$m profile is closer to
the much broader atomic gas component. Could dust associated with the HI
distribution be responsible of the emission at $200\mu$m? A model was
produced with two dust disks. As said earlier, a disk with 
$\alpha_{\mathrm{d}}=\alpha_\star$ can describe dust associated with 
the molecular gas. As for the dust associated to the atomic gas, we
chose a disk with a flat radial distribution up to 2.5$\alpha_\star$,
then falling off exponentially with a scalelength of 3 $\alpha_\star$.
This mimics the observed column density of HI \citep{TacconiApJ1986}.

Using the relation between extinction and gas column density of 
\citet{BohlinApJ1978} and the column density measured by 
\citet{TacconiApJ1986}, we derived an optical depth 
$\tau_\mathrm{V}\approx0.5$ for the broader dust disk. For the dust
disk associated with the molecular phase we adopted $\tau_\mathrm{V}=5$,
as in the previous standard disks. Using the observed H$_2$ column density, 
a value larger at least by a factor of two would have been derived.
However, too much energy would have been absorbed from dust (see, for
instance, the SED of the $\tau_\mathrm{V}=10$ standard model in
Fig.~\ref{sed_standard}). The discrepancy between the value derived from
the gas column density and that needed to produce the right amount of
absorption may be due to the smaller effective absorption in a clumpy 
structure of dust. Clumping is discussed in Sect.~\ref{discussion}. 

The SED of this
new model is shown in Fig.~\ref{sed_ext}. The temperature distribution is 
presented in the bottom panel of Fig.~\ref{fig_temp_ext}. Despite the colder 
temperature of the dust associated to HI, the behaviour of the FIR radiation 
in the region where scalelengths are measured is dominated by the dust disk 
associated with H$_2$\footnote{By extending the dust disk further, the
FIR scalelength can became as large as that observed. This happens, for
example, for a single HI-like dust distribution. However, optically thick 
models ($\tau_\mathrm{V}=2$ for the HI-like distribution) are still needed. 
Such models would have a dust mass larger than what implied assuming the 
canonical gas-to-dust mass ratio (Sect.~\ref{discussion}).}.
The derived scalelengths are similar to those of an extended model 
($\alpha_{\mathrm{d}}=1.5\alpha_\star$) with a single disk.

In conclusion, the optically thick regime is required to match the observed
FIR SED, both in models with standard and extended dust distribution.
FIR scalelengths are larger in extended models with respect to standard
models. The 100$\mu$m scalelength is very similar to that derived on
IRAS images, but the 200$\mu$m scalelength is always smaller than that
measured by ISO.  It is not possible to produce B/200$\mu$m scalelength 
ratios as small as in \citet{AltonA&A1998}, for any of the models presented 
here.

\section{Discussion}
\label{discussion}

As shown in the previous sections, models with a central face-on optical 
depth $\tau_V \sim 5$ are necessary to explain the SED observed in the FIR 
for NGC~6946.
A high optical depth through the central regions of the galaxy has also
been found by \citet{EngargiolaApJS1991} and \citet{DevereuxAJ1993}. 
\citet{EvansThesis1992} and \citet{TrewhellaThesis1998} apply the 
energy balance method to the stellar and dust emission of NGC~6946, 
using a TRIPLEX model, i.e. an analytical approximation for the radiative 
transfer, neglecting scattering, in a standard mode \citep{DisneyMNRAS1989}.
They both derived high optical depths for the disk, using the data 
inside the half light radius. \citet{EvansThesis1992} measured
$\tau_V=6-7$, while \citet{TrewhellaThesis1998} $\tau_V=4\pm1$.
A high optical depth is also suggested by high-resolution sub-mm
SCUBA images: the diffuse inter-arm emission in the NE spiral arms at a 
distance of 2$\arcmin$ ($\approx\alpha_\star$) is compatible with $\tau_V=2.2$
\citep{BianchiA&A2000}. \citet{XuA&A1995} carried out an energy balance on 
a sample of 134 nearby spirals with available UV, B and IRAS fluxes.
They derived a mean optical depth $\tau_B=0.60$. A direct comparison between 
their result and the optical depth needed by our model to explain the FIR 
output may lead to the wrong conclusion that we have overestimated the amount 
of absorbed energy. However, the mean ratio of bolometric
luminosity absorbed by dust for their sample ($\sim$ 1/3) is similar
to that predicted by our models for NGC~6946. The difference 
depend on their adopted geometry, a plane-parallel homogeneous model
for dust and stars.  If the dust is associated with the dominant 
gas component, like observations suggest, our choice of an exponential 
distribution seems more appropriate.
For the same optical depth, their model results much more effective in
extinguishing radiation. For example, in the UV, \citeauthor{XuA&A1995} 
adopt dust and stellar distributions of the same thickness 
\citep[a {\em slab} model][]{DisneyMNRAS1989} and isotropic scattering
with albedo $\omega$=0.18. By opportunely setting scalelengths and 
truncations, the BFG model can produce results for a slab. At their UV
reference wavelength $\lambda=2030$\AA, 60\% of the radiation is
absorbed by dust, for their adopted UV extinction law. Because of their
low albedo, scattering does not play a relevant role in the radiative
transfer. In our corresponding UV4 band, the $\tau_V=5$ standard model 
absorbs only 50\% of the radiation. In the optical, they adopt a {\em sandwich} 
model, with the thickness of the stellar distribution twice that for dust.
At $\lambda=4400$\AA, their model absorbs 20\% of the radiation, while
the $\tau_V=5$ standard model 40\%. 

In our simulations most of the radiation is absorbed in the Optical-NIR 
spectral range (60\% for light at $\lambda>4000$\AA). This was alredy
noted for NGC~6946 by \citet{TrewhellaThesis1998}. \citet{XuA&A1995}
reach opposite conclusions, with the non-ionising UV (912\AA
$<\lambda<$3650\AA) contributing to 60$\pm 9$\% of the absorbed radiation.
This is in part due to the model differences we have already 
discussed. \citeauthor{XuA&A1995} adopted a different thickness for the
UV stellar distribution to simulate the thinner distribution of younger
stars. We tried a similar approach by adopting the same scaleheight
for dust and star at smaller $\lambda$ \citep{BianchiThesis1999}, 
but the fraction of radiation absorbed in the UV increased only by a 5\%.
Furthermore, the SED observed in NGC~6946, with a peak in the 
NIR, may be different from the mean characteristic of the 
\citeauthor{XuA&A1995} sample. The UV selected sample may be biased
toward bright UV galaxies, although this hypothesis 
is dismissed in \citet{BuatA&A1996}.

Another major difference between our model and that of \citet{XuA&A1995}
is in the ionising UV longward of 912\AA . In their work the absorption of 
Lyman continuum photons contributes as much as 20$\pm 1$\% to the total FIR 
emission in a sample of 23 late type galaxies. Our omission of this
spectral range from the radiative transfer, due to the lack of
empyrical data, may result in an overestimate of the optical depth.
To test our approximation, we derived the ionising
UV from H$\alpha$  observations of NGC 6946 \citep{KennicuttAJ1983}, assuming 
standard ionisation conditions as in HII regions \citep{LequeuxProc1980}. 
After correcting for the [NII] contamination (25\% for spirals), Galactic 
extinction ($A_{\mathrm{H}\alpha}\approx 1$) and 
internal extinction (30\% for the R-band, where the H$\alpha$ line is 
located, for a standard $\tau_V=5$ model), we derived an intrinsic 
unextinguished H$\alpha$ flux,
\begin{equation}
f(\mbox{H$\alpha$})=9.1\; 10^{-11} \mbox{erg cm$^{-2}$ s$^{-1}$}.
\end{equation}
Following \citet{XuA&A1995}, the Lyman continuum flux can be derived as
\begin{equation}
f(\mbox{Lyc})=33.9 f(\mbox{H$\alpha$})
=3.1\;10^{-9} \mbox{erg cm$^{-2}$ s$^{-1}$},
\end{equation}
of which 75\% is assumed to be absorbed by gas and converted into emission 
lines at larger wavelengths 
\citep[see also][]{MezgerA&A1978,DeGioiaEastwoodApJ1992}.
Allowing the remaining 25\% to be entirely absorbed by dust and adopting
a lower limit MIR correction of $\approx 70$\% (Sect.~\ref{mircorre}), the 
total FIR luminosity originating from the absorption of ionising photons is
\begin{equation}
L^{\mathrm{FIR}}(\mbox{Lyc})=2.2 \;10^{8} \mbox{L}_\odot,
\end{equation}
for the assumed distance of 5.5 Mpc. This correspond to only $\approx$
2\% of the total FIR luminosity emitted by the standard $\tau_V$=5
model.  As a comparison, the contribution to the FIR from the EUV band for the
same model is 3.6\%. The ionising UV contribution is similar for all the
models that roughly provide the same amount of FIR energy as that
observed. Therefore, it is justified to omit the ionising radiation in
the model. The approach of \citet{XuA&A1995} is different from the one
presented here. Their UV contribution includes direct absorption 
of Lyc photons and indirect (via emission lines), while in the
present model the absorption of emission line photons is taken care
of in the spectral band where the emission occurs (e.g. in the R-band
for the H$\alpha$ line), the total contribution of re-combination is summed 
up to the stellar SED for each band.
Nevertheless, their ratio between Lyc emission and total absorbed energy 
is similar to the one derived here.

A wrong MIR correction could also be responsible for an underestimation
of the FIR emission in our model.
The fraction of absorbed energy that goes into MIR emission depends 
essentially on the absorption of light from the short wavelength spectrum, 
since the absorption efficiency of small grains responsible for
non-equilibrium processes is higher in the UV (Sect.~\ref{mircorre}).
For the model presented here, with a dust scaleheight smaller than the 
stellar one, the 
amount of energy absorbed from UV bands does not increase very much with the 
optical depth (The {\em saturation effect}, \citealt{BianchiApJ1996}). 
The MIR corrections for these models are therefore quite constant, 
$\sim$32\% of the total absorbed energy being re-emitted in the MIR. 
Even for models with higher efficiency in extinguishing radiation, like
the optically thick model with a thicker dust distribution, the fraction of
the total energy emitted in the MIR is not very different from this value 
\citep{BianchiThesis1999}.

For the local ISRF the contribution of small grain emission to the 60 $\mu$m 
IRAS band is $\sim$62\%, while at 100 $\mu$m it is only 14\% and at 200 $\mu$m
4\% \citep{DesertA&A1990}. Therefore, the fraction of energy emitted in non 
equilibrium heating can be roughly estimated by measuring the MIR emission 
shortward of 60 $\mu$m. After integrating a continuous SED interpolated 
from the data points in Table~\ref{tab_sed}, the MIR energy is derived
to be 34\% of the total infrared energy emitted by dust. 
The value derived from observation is very close to the model one.
This justifies the use of the \citet{DesertA&A1990} dust model as
described in Sect.~\ref{mircorre}. It is interesting to note that the 
infrared galactic spectrum used in the \citet{DesertA&A1990} model is 
different from the one of NGC~6946.  As an example, the ratio between 
fluxes at 60 $\mu$m and 100 $\mu$m is 0.2, while it is 0.5 from the 
NGC~6946 data.  This does not necessarily mean that the dust model of
\citet{DesertA&A1990} cannot be applied to NGC~6946. The different ratio
could be due to different heating conditions in the local interstellar
radiation field, with respect to the mean radiation field of NGC~6946.
Larger ratios between 60 $\mu$m and 100 $\mu$m can be derived 
from the \citet{DesertA&A1990} model when the ISRF is larger than the local.
As already said, emission from small grains can contribute to part of
the observed FIR flux. However, for a wide range of heating conditions, the 
contribution close to the peak of FIR emission is minimal 
\citet{DesertA&A1990}. Since modelled and observed fluxes are compared 
in this spectral range, the assumption that all FIR emission occurs at 
thermal equilibrium does not affect sensibly our results.

The position of the peak emission in the modelled FIR SED clearly shows
that the dust temperature in the simulations is not severely different
from the real one. Recently, \citet{StickelA&A2000} extracted high
signal-to-noise sources from the ISOPHOT Serendipity Survey. Fluxes at 170 
$\mu$m were measured for 115 sources having a galaxy association, most
of which are spirals, and integrated with 100$\mu$m IRAS data. The distribution
of the ratio of 170$\mu$m and 100$\mu$ fluxes is quite well confined,
with half of the galaxies having a ratio between 1 and 1.5. For the
emissivity adopted here, this translates in a colour temperature in
the range 23-28 K. Similar colour temperatures (24-26K) can be derived
from the ratios of the total fluxes at 100 and 200$\mu$m, in any of our
models. Using emissivities derived from Galactic emission and exinction,
\citet{BianchiA&A1999} found that the value of the dust mass for a small 
sample of galaxies does not depend dramatically on the assumed
spectral behaviour of the emissivity. This can be easily interpred if dust 
temperatures in external galaxies are similar to those in the Milky Way.
Because of these similarities, and of the small dependence of the
temperature on the assumed model, it is not too audacious to compare 
values measured on the present models with the high quality observations
of the Galaxy.

\begin{figure}
\resizebox{\hsize}{!}{
\includegraphics{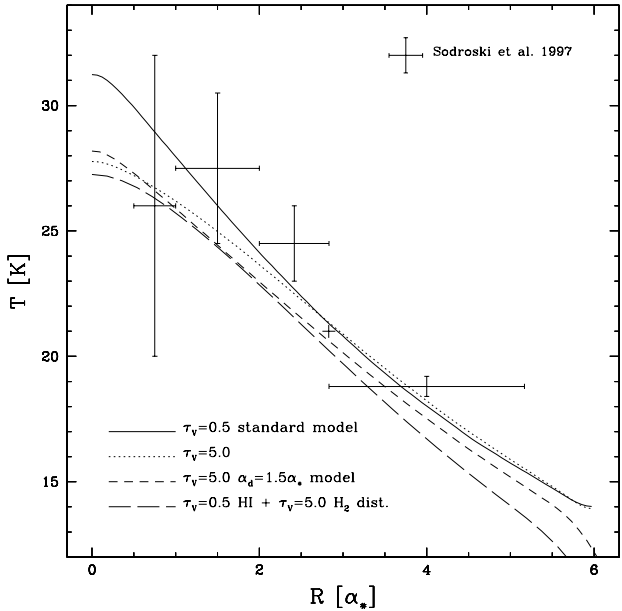}}
\caption{Temperature along the galactic plane as a function of the
galactocentric distance for four different models: the
$\tau_\mathrm{V}=0.5$ and $\tau_\mathrm{V}=5$ standard models, a 
$\tau_\mathrm{V}=5$ $\alpha_\mathrm{d}=1.5\alpha_\star$ model and 
the model with a dust disk associated with each of the two gas phases
(Sect.~\ref{results}). Data points show the radial gradient of
temperature in the Galaxy and are derived from \citet{SodroskiApJ1997} 
as described in the text.  The vertical error bar 
represent the error in the temperature determination, while the horizontal 
the radial range for which the temperature has been measured.
The cross marks the temperature of 21K at the Sun distance from
the galactic centre \citep{BianchiA&A1999}.}
\label{sodcon}
\end{figure}

Dust temperatures at high latitude are quite constant in the Galaxy
\citep{ReachApJ1995,LagacheA&A1998,SchlegelApJ1998}.
For the  emissivity adopted here, a temperature of $\approx$21K can
be derived \citep{BianchiA&A1999}. Remarkably, nearly all the models
present a similar temperature at a galactocentric distance of
3$\alpha_\star$ (the Sun position, for $R_\odot$=8.5kpc and a Galactic
scalelength of 3kpc). \citet{SodroskiApJ1997} decompose the Galactic FIR 
emission observed by DIRBE into three components, associated with the 
atomic, molecular and the ionised gas phases, and derive the temperature 
for four annulii at different galactocentric distance. Fig.~\ref{sodcon} 
shows the temperature of the dust associated with the atomic gas 
(supposedly heated mainly by the diffuse ISRF) for each annulus. Data 
have been scaled to a Galactic scalelength $\alpha_\star=3$ kpc and 
corrected for the emissivity law used in this work. In Fig.~\ref{sodcon}
we also plot the temperature radial profile for four representative
models: the $\tau_\mathrm{V}=0.5$ and $\tau_\mathrm{V}=5$ standard models; 
the $\tau_\mathrm{V}=5$ $\alpha_\mathrm{d}=1.5\alpha_\star$ model; and the 
model with a dust disk associated with each of the two gas phases. Clearly,
all the simulations presented here have a temperature gradient compatible 
with the Galactic one.

Because of the lack of high resolution observations in the FIR, the
derivation of  temperature gradients in external galaxies is more difficult.
\citet{DaviesMNRAS1999} derived the temperatures 
from 100$\mu$m IRAS and 200$\mu$m ISO fluxes at two different position 
on NGC~6946, in the centre and on the disk at 3 arcmin from the centre.
After smoothing for the ISO resolution, our models are compatible with
those observations \citep{BianchiThesis1999}.

For any model, the FIR scalelengths increase with the wavelength of
emission, the $\alpha_\mathrm{d}/\alpha_\star$ ratio and, although
slightly, with the optical depth. Therefore, larger FIR scalelengths can be 
found in optically thick extended models. For the $\tau_\mathrm{V}=5$ extended
model, that provides a good fit to the SED, the 100$\mu$m scalelenght
is very close to the value derived from IRAS images. The $200\mu\mathrm{m}$ 
scalelength is larger than scalelength at 100$\mu$m. However, it is 
30\% smaller than that observed by ISO. It has not been possible to find
a model able to reproduce the large 200$\mu$m scalelength measured by
\citet{AltonA&A1998} in a sample of seven galaxies including NGC~6946.
The B-band ratio is close to that observed for the $\tau_\mathrm{V}=5$
extended and standard models.  \citeauthor{AltonA&A1998} derived a ratio 0.9 
between the B and 200$\mu$m scalelengths. Because of the dependence of the 
B-band scalelength on $\tau_\mathrm{V}$, similar ratios can be obtained
only for optically thin extended models. Such models fail to predict
both the SED and the absolute values of the FIR scalelengths. A possible
reason for the discrepancy between observations and models may reside in
the transition effects of the ISOPHOT detectors. Memory effects during
the scanning of a bright source affects the P32 mapping mode.
\citeauthor{AltonA&A1998} 
checked for this problem by conducting the scalelength analysis on both
in-scan and cross-scan directions and concluded that the effects were
negligible. However, a proper analysis requires a knowledge of the
underlying light distribution and  a description of the transient
effects, which are still poorly understood. The P32 mapping mode is still
not scientifically validated \citep{KlaasRep2000}.

We have carried out a 
few test to check for the influence of different stellar distributions, like 
including a small bulge appropriate for a late type galaxy or having
a smaller vertical scalelength for younger stars \citep{BianchiThesis1999}.
The result were not significantly different from those presented here. 
A basic assumption of all our models is that the intrinsic (not
extinguished) radial stellar scalelength is the same at any wavelength. 
According to this view, the larger scalelengths observed in the optical are 
due to extinction, rather than to intrinsic color gradients. As outlined in 
Sect.~\ref{stardisk}, the problem is still very debated. A simple test can 
be conducted here to assess the influence of scalelength variations with
$\lambda$. Let's assume that the measured B-band scalelength
(1.3$\alpha_\star$) is more representative of the intrinsic radial variation 
of the stellar distribution. By substituting $\alpha_\star$ with
1.3$\alpha_\star$, the results of Table~\ref{tab_scale}show that
standard or extended models with $\tau_\mathrm{V}=0.5$ or 1 have an 
{\em observed} (after smoothing) B-band scalelength quite similar to the
one of \citet{AltonA&A1998}, the difference between observed and
intrinsic scalelength due to the smoothing process, rather than to extinction.
Since the modelled B/200$\mu$m scalelength ratio remains the same,
these optically thin cases may have a 200$\mu$m scalelength close to the
observed. However, the fraction of energy absorbed in these models will
remain the same as well. Therefore, the optically thin models will still be 
unable to produce the required energy output in the FIR. Furthermore,
most of the energy is absorbed in the NIR, where the peak of stellar 
emission occurs. In this spectral range  our K-band based $\alpha_\star$ is 
surely a better description for the galaxy radial scalelength.

We also produced a model with two dust distributions, a $\tau_\mathrm{V}=5$ 
standard disk associated with the molecular gas dominant in the central part 
of NGC~6946, and a $\tau_\mathrm{V}=0.5$ disk associated with the broader 
distribution of atomic gas. Even with the presence of such an extended 
distribution, the double disk model is dominated by the optically thick disk 
associated to the H$_2$ necessary for the FIR emission. For what concerns
FIR scalelengths, the same results as for the $\tau_\mathrm{V}=5$ extended 
model are obtained.  It is interesting to note that for the parameters adopted 
here for the two dust disks and assuming the atomic + molecular gas mass of 
Table~\ref{tab_basic}, the gas-to-dust mass ratio is 180, close to the 
Galactic value of 160 \citep{SodroskiApJ1994}.
Similar values can be retrieved from all the optically thick models.

The high optical depth of the models contrasts with the recent
determination of optical depth in edge-on spirals by
\citet{XilourisSub1998}. Using a sample of seven edge-on galaxies, they
find a mean central face-on optical depth $\tau_\mathrm{V} \approx 0.5$.
The higher opacity of NGC~6946 may be a result of the galaxy being very
gas-rich \citep{TacconiApJ1990}; or it may be due to clumping of the
ISM, affecting in a different way FIR and optical determinations of the
optical depth. While FIR observations detect all of the dust
(at least when the temperature of the clump and inter-clump medium are
similar), optical observations may be affected preferentially by
the extinction of the smoother, lower density (and optical depth)
inter-clump medium.

To our knowledge, only two works in literature discuss the effects of 
clumping within the framework of a radiative transfer model for spiral
galaxies. \citet{KuchinskiAJ1998} distribute clumps in the dust disk
assuming a constant filling factor and a value for the ratio
between densities in clumps and in the nearby smooth medium.
The other model is by \citet{BianchiSub1999}, based on the BFG
radiative transfer code, as the present work. \citet{BianchiSub1999}
modelled the clump distribution in a similar way to the distribution of 
molecular gas in the Galaxy. The clump properties were derived from those 
observed in Giant Molecular Clouds. A comparison with the 
\citet{KuchinskiAJ1998} results indicates a strong dependence of the 
observed brightness profiles on the detailed internal and spatial 
distribution properties of clumps. This makes the interpretation of the data 
very difficult.
For the same dust mass, it is found that a clumpy dust medium has lower
extinction. However, when a fraction of the stellar radiation is allowed
to be emitted from inside the clumps, as for Giant
Molecular Clouds hosting star-formation, extinction increases and
can reach higher values than those for homogeneous models.

Predictions of the influence of clumping on the FIR models are not easy.
If the dust component associated with the H$_2$ is clumped, it could be
responsible for most of the FIR emission, when embedded stellar emission
is considered. A diffuse component associated with the HI would be
responsible for a reduced $\tau_\mathrm{V}$ derived from the analysis
of edge-on galaxies. For the models of \citet{BianchiSub1999}, it is unlikely
that a clumpy distribution with a large dust mass (corresponding to an
optically thick smooth distribution) has an optically thin apparent
$\tau_\mathrm{V}$ as measured by \citet{XilourisSub1998}.
However, the models of \citet{BianchiSub1999} are based on the Galaxy,
while clumping based on the different distributions of atomic and molecular 
gas in NGC~6946 may have a different behaviour. Clumping will also
affect the spatial distribution of the FIR emission. Dust in clumps
is shielded from the ISRF and heated to lower temperature. Depending on
the distribution of cold dust clouds, a broader 200$\mu$m emission could
be produced. 

Two recent models include clumping of dust and embedded stellar emission
to describe the radiative transfer and FIR emission of NGC~6946. However, 
the complexity of the models prevents an isolation of the effect of the 
dust distribution on the FIR heating. \citet{SilvaApJprep1998} fit the
observed optical and FIR SED of NGC~6946 with their galactic photometric
evolution model, adopting a simplified treatment for the radiative
transfer in the diffuse medium and a separate model for dust in clumps.
They find that nearly half of the dust emission comes from within molecular 
clouds. The adopted distributions for smooth dust and stars are different 
from those in Sect.~\ref{results}: the radial scalelengths are
the same for both dust and stars, as well as the vertical scalelengths;
the radial scalelength is twice the one adopted here; the disk is
thicker by a factor of two.
Since we use the same SED and their model is optically thin to
B-band radiation (only 10\% of radiation absorbed), most of the FIR
emission must come from absorption of UV photons in the molecular cloud
distribution. A distribution of molecular clouds compatible with the
galactic gravitational potential, together with a smooth phase modelled
on the atomic gas, is used by \citet{SautyA&A1998} to study the
FIR emission in NGC~6946. The radiative transfer is carried out using
the Monte Carlo technique only for the UV radiation emitted in star-forming
regions within the molecular clouds. The ISRF at $\lambda>2000$\AA\ is
derived from a R-band map of the galaxy, scaled on the Galactic local
ISRF. As in our case, the observed FIR emission can be explained only
with substantial extinction. However, they find that emission at 
$\lambda<2000$\AA\ contributes 72\% of the total FIR, while in the present 
work energy absorbed from optical radiation is dominant.

Therefore, according to the models of \citet{SilvaApJprep1998} and
\citet{SautyA&A1998}, the dominant contribution to the dust heating
comes from young stars. If this is true, the recent star-formation rate
of a galaxy could be measured using FIR fluxes. These are more readily 
available than other tracers of star-formation \citep{DevereuxApJ1991}.
In a series of papers, \citet[][and references therein]{DevereuxAJ1993}
compared FIR and H$\alpha$ fluxes and suggested that the FIR luminosity
is dominated by warm dust absorbing radiation from OB stars. However,
this result is debated. \citet{WalterbosApJ1996} modelled the FIR
emission in spiral galaxies, deriving the diffuse ISRF from optical profiles 
and estimating the dust column density from the atomic gas. For a sample of
20 galaxies, they found that dust heated by a diffuse ISRF can account
for, on average, half of the observed IRAS fluxes. Using large nearby objects
it is possible to decompose the FIR emission into the different sources of 
heating. \citet{XuApJ1996a} measure the ratio between the IRAS 60$\mu$m
and H$\alpha$ fluxes from bright FIR-resolved sources in M31. Using the
total H$\alpha$ luminosity and the \citet{DesertA&A1990} dust model,
they extrapolate the fraction of the total luminosity emitted by dust that is
associated with HII regions and star-formation. A value of 30$\pm$14\% only
is found. A decomposition of the Galactic flux observed by COBE at
140$\mu$m and 240$\mu$m shows that most of the dust emission ($\sim$70\%) 
arises from dust associated with the atomic gas \citep{SodroskiApJ1994},
with temperature gradients compatible with a diffuse ISRF heating.
Only 20\% of the FIR is emitted by dust associated with the molecular
component and 10\% is due to hot dust associated with the HII phase.

The smooth models presented here attempt to explain the FIR emission without
invoking dust heated in star-forming regions. Under the assumption of a
diffuse stellar emission and dust, we find that radiation at 
$\lambda < 4000$\AA\ contributes $\approx$ 30\% to the total FIR, this
including the ionising UV. This percentage will surely increase when a clumpy 
dust structure is adopted. The high optical depths necessary to produce 
the observed energy may be an overestimation because we neglect 
clumpy hot dust. However, without a proper model for the radiative
transfer and dust emission in a clumpy medium, it is not easy to 
understand if the results shown here are severely biased by the 
assumption of smooth distributions.

Finally, we tried to model the FIR emission with an extra dust component, a 
spherical halo. The presence of dust in the halo of normal, non-active,
spirals is predicted as a result of the imbalance between the radiation 
pressure and the galactic gravitational force \citep{DaviesMNRAS1998}. 
A halo of cold gas clouds, able to explain part of the dark matter in
spiral galaxies, may be stabilized by the presence of dust grains
\citep{GerhardApJ1996}. Unfortunately, it is difficult to obtain information 
about a putative dusty halo. Because such a distribution would act as a
screen for the galactic disk, the radiative transfer fits of 
\citet{XilourisSub1998} are unable to detect it. Analysing the
difference in colours of background objects between fields
at different distances from the centre of two nearby galaxies,
\citet{ZaritskyAJ1994} find a B-I colour excess in the inner fields.
He derived a dust halo scalelength of 31$\pm$8 kpc, although the
measured colour difference is only 2$\sigma$ of the statistical noise.

Because of the high uncertainty of the halo parameters, we simply used a
spherical homogeneous dust halo, together with a standard disk. The halo
has the same radial dimension as the dust and stellar disks,
i.e. 6$\alpha_\star$. Among the models we tried, the case in which
$\tau_\mathrm{V}=1$ for both  halo and disk provides a reasonable fit to
the FIR emitted energy and scalelengths \citep{BianchiThesis1999}.
However, when the galaxy is seen edge-on, such a dust distribution would be 
easily observable, within the resolution and sensitivity of available
FIR observations. \citet{AltonMNRAS1998} studied the FIR emission in 24
edge-on galaxies, including starburst and quiescent objects, using HiRes
IRAS images. None of the object was found to be resolved along the minor axis.

\section{Summary and conclusions}

We have described in this paper a model for dust emission in spiral galaxies, 
based on the Monte Carlo radiative transfer code  of \citet*{BianchiApJ1996}.
For each relative star-dust geometry and dust optical depth, the radiative 
transfer code is carried out for 17 different photometric bands, covering 
the spectral range of stellar emission. In the application to NGC~6946
presented here, we have chosen a model stellar SED and dust distribution
that produces the same fluxes as those observed in the UV-Optical-NIR images.
The code also produces a map of the total energy
absorbed by dust. For each position inside the galaxy, dust is heated
by an ISRF consistent with the radiative transfer itself, without any other
assumption. The dust temperature is computed from the absorbed energy, 
using the emissivity derived by \citet{BianchiA&A1999} for Galactic dust 
and correcting for the contribution of non-thermal equilibrium processes 
to the emission. Hence, FIR maps can be easily obtained for any wavelength,
integrating along a specific line of sight. 

The model optical and FIR scalelengths and the SED have been compared to 
NGC~6946 observations. Several models have been explored. It is found
that optically thick dust disks ($\tau_\mathrm{V}\approx$5) are needed to
match the observed FIR output. Approximately one third of the total
stellar radiation is absorbed by dust in this case. 
The temperature distributions are quite similar, for any of the dust disk 
models. Temperature values in the models are comparable with those observed
in our Galaxy and other spirals.

We have compared the modelled FIR scalelengths with the observations of
\citet{AltonA&A1998}. Extended dust disk model with
$\alpha_{\mathrm{d}}=1.5\alpha_\star$ \citep{DaviesMNRAS1997,XilourisSub1998} 
have larger FIR scalelengths than standard models with 
$\alpha_{\mathrm{d}}=\alpha_\star$. For optically thick cases, the
100$\mu$m scalelength is close to the value derived on IRAS images. 
\citet{AltonA&A1998} found that the B-band scalelength of NGC~6946 is smaller 
than the 200$\mu$m one by a factor 0.9 (0.8 for a sample of seven galaxies).
We have not been able to reproduce the large FIR scalelengths 
measured on 200$\mu$m ISO data. In the required optically thick regime, 
the scalelength ratio B/200$\mu$m is always larger than observed.
Smaller ratios can be obtained only in optically thin cases, but the
absolute values for the scalelengths are smaller.
The results are not improved if two dust disks modelled on the gas
distribution are used. The behaviour of the model is dominated by the
standard optically thick disk associated with the molecular component, 
rather than the very extended dust distribution associated with HI.
A spherical dust halo could produce results closed to those observed, but 
would also be easily detected in currently available FIR observations, 
which is not.

The high optical depth found for NGC~6946 contrasts with recent determinations
on edge-on spiral galaxies \citet{XilourisSub1998}. This may be a result of 
our assumption of a smooth distribution for stars and dust. The inclusion of 
clumping in a proper model of radiative transfer and FIR emission is therefore 
desirable. However, the heavy dependence of clumping on the assumed model 
makes the modelling more complicate. Future high resolution and sensitivity 
instrumentation will therefore be essential to define the dust distribution 
and limit the number of parameters in the model. Clumping may also be 
responsible for the discrepancy between the observed and modelled scalelengths 
at 200 $\mu$m, if a diffuse component of cold dust clumps shielded from
the ISRF is present.  On the other hand, it is necessary to remind here
that the results of \citet{AltonA&A1998} at 200$\mu$m are based on data which 
is not yet scientifically validated. Again, future FIR instrumentation or a 
set of validated data will help to asses if the large scalelengths are an
artefact of the ISOPHOT detector transients or if they are the genuine
results of a temperature distribution different from the one we have
derived here.

The main purpose of the paper was the presentation of the model itself. 
A test has been conducted on a well studied galaxy. However, NGC~6946
is very gas rich \citep{TacconiApJ1990} and its characteristics may be
different from those of a 'mean' galaxy. Optical and FIR data are
available for several galaxies. A future paper will be devoted to their
analysis and more general conclusions about the dust distribution and 
extinction will be drawn.

\begin{acknowledgements}
The work of this paper has benefitted from comments and discussion with
many people.  Among them we wish to remember M. Trewhella, Z. Morshidi, 
R. Smith, A.  Kambas, J. Haynes, Rh. Evans, Rh. Morris, A. Whitworth, 
A. Ferrara, S. Kitsionas, P. Gladwin, N. Francis and an anonymous referee. 
S.B. acknowledges a PhD studentship from the Department of Physics and 
Astronomy at Cardiff University.
\end{acknowledgements}

\end{document}